\documentclass[twocolumn]{aastex631}
\usepackage{amsmath}

\begin{document}

\title{WASP-121\,b's transmission spectrum observed with JWST/NIRSpec G395H reveals thermal dissociation and SiO in the atmosphere}

\author[0009-0007-9356-8576]{Cyril Gapp\email{gapp@mpia.de}}
\affiliation{Max-Planck-Institut f\"{u}r Astronomie, K\"{o}nigstuhl 17, D-69117 Heidelberg, Germany}
\affiliation{Department of Physics and Astronomy, Heidelberg University, Im Neuenheimer Feld 226,
D-69120 Heidelberg, Germany}

\author[0000-0001-5442-1300]{Thomas M. Evans-Soma}
\affiliation{School of Information and Physical Sciences, University of Newcastle, Callaghan, NSW, Australia}
\affiliation{Max-Planck-Institut f\"{u}r Astronomie, K\"{o}nigstuhl 17, D-69117 Heidelberg, Germany}

\author[0000-0003-3726-5419]{Joanna K. Barstow}
\affiliation{School of Physical Sciences, The Open University, Walton Hall, Milton Keynes, MK7 6AA, UK}

\author[0000-0003-3667-8633]{Joshua D. Lothringer}
\affiliation{Space Telescope Science Institute, 3700 San Martin Drive, Baltimore, MD 21218, USA}
\affiliation{Department of Physics, Utah Valley University, 800 West University Parkway, Orem, UT 84058, USA}

\author[0000-0001-6050-7645]{David K. Sing}
\affiliation{Department of Earth \& Planetary Sciences, Johns Hopkins University, Baltimore, MD, USA}
\affiliation{Department of Physics \& Astronomy, Johns Hopkins University, Baltimore, MD, USA}

\author[0009-0005-9163-0095]{Djemma Ruseva}
\affiliation{Max-Planck-Institut f\"{u}r Astronomie, K\"{o}nigstuhl 17, D-69117 Heidelberg, Germany}
\affiliation{University of St Andrews, North Haugh, St Andrews, KY16 9SS, UK}

\author[0000-0003-0973-8426]{Eva-Maria Ahrer}
\affiliation{Max-Planck-Institut f\"{u}r Astronomie, K\"{o}nigstuhl 17, D-69117 Heidelberg, Germany}

\author[0000-0002-8515-7204]{Jayesh M. Goyal}
\affiliation{School of Earth and Planetary Sciences (SEPS), National Institute of Science Education and Research (NISER), Jatani, India}

\author[0000-0002-4997-0847]{Duncan Christie}
\affiliation{Max-Planck-Institut f\"{u}r Astronomie, K\"{o}nigstuhl 17, D-69117 Heidelberg, Germany}

\author[0000-0003-0514-1147]{Laura Kreidberg}
\affiliation{Max-Planck-Institut f\"{u}r Astronomie, K\"{o}nigstuhl 17, D-69117 Heidelberg, Germany}

\author[0000-0001-6707-4563]{Nathan J. Mayne}
\affiliation{Department of Physics and Astronomy, Faculty of Environment, Science and Economy, University of Exeter, Exeter EX4 4QL, UK}

\begin{abstract}
WASP-121\,b has been established as a benchmark ultrahot Jupiter, serving as a laboratory for the atmospheric chemistry and dynamics of strongly irradiated extrasolar gas giants. Here, we present and analyze WASP-121\,b's transmission spectrum observed with NIRSpec G395H on board the James Webb Space Telescope and find evidence for the thermal dissociation of H$_2$O and H$_2$ on the planet's permanent dayside. Additionally, we detect SiO at a statistical significance of $5.2\sigma$ which is compatible with chemical equilibrium in the atmosphere. Constraining the abundance of SiO and abundance ratios between silicon and volatile atoms in WASP-121\,b's atmosphere could help discriminate between possible migration histories of the planet. The three-dimensional nature of thermal dissociation on WASP-121\,b's dayside and of recombination on its nightside, however, poses a challenge to constraining molecular abundances and elemental abundance ratios from the transmission spectrum. To account for this, we implemented an atmospheric model in the \texttt{NEMESIS} framework that splits the planet's atmosphere into dayside and nightside. A retrieval applying our atmospheric model to WASP-121\,b's transmission spectrum favors a higher H$_2$O abundance on the nightside than on the dayside, demonstrating the impact of hemispheric heterogeneity when attempting to constrain WASP-121\,b's bulk H$_2$O inventory.
\end{abstract}

\keywords{Exoplanet atmospheric composition (2021) --- Transmission spectroscopy(2133) --- Infrared spectroscopy (2285)}

\section{Introduction} \label{sec:introduction}
WASP-121\,b is a gas giant exoplanet with a mass of $1.16\pm 0.07$ Jupiter masses and a radius of $1.75\pm 0.04$ Jupiter radii that orbits an F-type star with a period of $1.274925$ days \citep{bourrier20}. At its equilibrium temperature ($T_\text{eq}$) of about 2350\,K \citep{delrez16}, it is classified as an ultrahot Jupiter, because its dayside is hot enough to dissociate most molecules and ionize atoms, fundamentally changing the dynamical and spectroscopic properties of the atmosphere as compared to planets with $T_\text{eq}\lesssim 2000$\,K (see, e.g., \citealp{bell18,parmentier18,lothringer20,beltz21}). Since most chemical elements are expected to be vaporized in the extreme temperatures of ultrahot Jupiters (see, e.g., \citealp{kitzmann18,lothringer18}), WASP-121\,b offers the opportunity to constrain the abundances of its constituent species using atmospheric observations. This opens up a promising path to infer its formation history, because the amount of rocky material accreted during the planet's formation and eventual migration can be inferred from its refractory-to-volatile elemental abundance ratios (see, e.g., \citealp{lothringer21}). Measuring these ratios simultaneously with other elemental abundance ratios, such as the carbon-to-oxygen ratio (C/O), delivers leverage into constraining the planet's formation location in the protoplanetary disk.

As an exoplanet outstandingly suitable for atmospheric characterization, WASP-121\,b has been regularly observed using both ground-based and space-based observatories. Eclipse observations with the Hubble Space Telescope (HST) have revealed a thermal inversion on its permanent dayside \citep{Evans17,MikalEvans19,MikalEvans20}, implying the existence of short-wavelength absorbers in the atmosphere (see, e.g., \citealp{burrows08,fortney08}). This has been confirmed by WASP-121\,b's transmission spectrum from the near-ultraviolet (NUV) to the optical observed with HST which revealed a rise in transit depth toward shorter wavelengths in the NUV \citep{Evans18}. \cite{Evans18} were not able to identify any individual short-wavelength absorber conclusively using these observations; however, they suggested VO as a possible optical absorber, did not find evidence for TiO and noted that SH might have caused the observed NUV rise. Ground-based observations of WASP-121\,b obtained using HARPS have suggested the presence of neutral vanadium that could give rise to VO \citep{Hoeijmakers20}, while atomic titanium and TiO have been ruled out using high-resolution ESPRESSO spectra \citep{hoeijmakers24}, lending plausibility to the suggestion of \cite{Evans18}. \cite{lothringer22} argued that the observed NUV rises in WASP-121\,b ($T_\mathrm{eq}\sim 2350$\,K) and WASP-178\,b ($T_\mathrm{eq}\sim 2450$\,K) might be related to the absence of such a rise in HAT-P-41\,b ($T_\mathrm{eq}\sim 1950$\,K) via the condensation of silicate clouds in gas giants with equilibrium temperatures below $\sim 2100$\,K. In that picture, the NUV rises in WASP-121\,b and WASP-178\,b would not be created by SH, but instead by either SiO or Mg I and Fe II in tandem, while any silicon and magnesium in HAT-P-41\,b would be condensed into silicate clouds. Neither Mg I, Fe II or SiO could be identified unambiguously in either WASP-121\,b or WASP-178\,b, though, due to their overlapping opacities in the NUV. The definite identification of the absorber creating the NUV rise in WASP-121\,b has since been pending.

Since WASP-121\,b is a highly irradiated, tidally-locked planet, the atmospheric temperatures are dramatically higher on its dayside than on its nightside \citep{MikalEvans22,MikalEvans23}. This difference in temperatures probably drives chemical differences between the two hemispheres, because the dayside is hot enough to dissociate most molecules, including H$_2$, while the temperatures on the nightside are low enough for dissociated species to recombine. Indeed, phase-curve observations of WASP-121\,b obtained with HST have revealed stratospheric dayside temperatures high enough to dissociate H$_2$O, while the nightside temperatures are too low for the thermal dissociation of H$_2$O \citep{MikalEvans22}. In line with these phase-curve observations, high-resolution transit observations of WASP-121\,b using Gemini-S/IGRINS revealed that the observed H$_2$O absorption lines moved toward longer wavelengths with time \citep{wardenier24} which suggests a higher H$_2$O abundance on the nightside than on the dayside \citep{wardenier23}. In contrast, the observed CO absorption lines became increasingly blueshifted during the observed transits, indicating that CO, unlike H$_2$O, is abundant on the dayside. These findings are in agreement with predictions from general circulation models (GCMs) of a depletion of H$_2$O on WASP-121\,b's dayside due to thermal dissociation and a nearly globally uniform distribution of CO which is expected to be stable enough to avoid dissociation \citep{parmentier18}.

This dichotomy between the global distributions of H$_2$O and CO abundances probably does not only impact phase-curve and high-resolution observations of WASP-121\,b, but also low-resolution transmission spectra as the stellar light that filters through the planet's extended atmosphere during transits probes parts of both the dayside and the nightside (see, e.g., \citealp{caldas19}). WASP-121\,b's dayside is substantially inflated over the nightside due to its much higher temperatures and lower mean molecular weight caused by the thermal dissociation of H$_2$ into atomic hydrogen. Thus, the amplitude of absorption bands of any molecule that is abundant on the dayside, such as CO, will be increased, while the absorption amplitudes of molecules that are more abundant on the nightside, such as H$_2$O, will be decreased. Neglecting these hemispheric differences in a one-dimensional model that prescribes spherical symmetry can thus lead to an overestimation of the planet's CO abundance and an underestimation of the H$_2$O abundance, leading to a C/O that is biased to higher values \citep{pluriel20}.

\section{Observation and data reduction} \label{sec:observations}
In this work, we analyze a transit of WASP-121\,b observed as part of the phase-curve observations of James Webb Space Telescope (JWST) program GO-1729 (PI: Evans-Soma, co-PI: Kataria). The observation was carried out with the NIRSpec instrument employing G395H to obtain spectra between 2.54 and 5.15$\,\mu$m with a wavelength gap ranging from 3.72 to 3.82$\,\mu$m caused by the offset between NIRSpec's NRS1 and NRS2 detectors. As reported in \cite{MikalEvans23}, the telescope pointing and performance throughout the 37.8\,hr long phase-curve observation were very stable. A data reduction of the full phase-curve observation including the transit is presented by \cite{EvansSoma25} and here, we analyze the transit separately.

\subsection{Transit-only analysis}
We extracted the target spectrum for each integration of the observation from the respective images using the \texttt{FIREFLY} code \citep{rustamkulov22,rustamkulov23} as previously outlined in \cite{MikalEvans23}. To obtain WASP-121\,b's transmission spectrum, we analyzed the data from 3.5\,hr before to 3.5\,hr after the transit midtime, leaving about 2.05\,hr of both pre- and post-transit baseline for calibrating stellar fluxes and instrumental systematics in addition to the transit with a duration of about 2.9\,hr itself. Compared to a pure transit observation, our transit-only analysis of the phase-curve data has the advantage of the exposure beginning many hours prior to the transit, allowing plenty of time for preliminary settling systematics to subside.

For the analysis of the transit light curves we fit a model of the form
\begin{equation}
    M(c_0,c_1,c_2,\theta,t) = B(c_0,c_1,c_2,t) T(\theta,t), \label{eq:transit_model}
\end{equation}
to the data in each wavelength bin. Here, 
\begin{equation}
    B(c_0,c_1,c_2,t)=c_0 + c_1 t + c_2 t^2
\end{equation}
is a quadratic baseline incorporating both stellar and planetary fluxes as well as instrumental systematics and $T(\theta,t)$ is a transit model calculated using the \texttt{batman} package \citep{batman}. The parameters $\theta$ of the transit model we chose are two stellar limb darkening coefficients ($u_1$ and $u_2$), the planet's orbital period ($P$), eccentricity ($\epsilon$), inclination ($i$), semi-major axis relative to the star's radius ($a/R_*$), and radius relative to the star's radius ($R_p/R_*$); and the midtime of the transit ($T_\text{mid}$). For all models, we fixed WASP-121\,b's argument of periapsis to $90^\circ$, its orbital eccentricity to $0$, and its orbital period to 1.27492504 days, as was reported by \cite{bourrier20}.

\begin{figure*}[htbp!]
	\centering
	\includegraphics[width=\hsize]{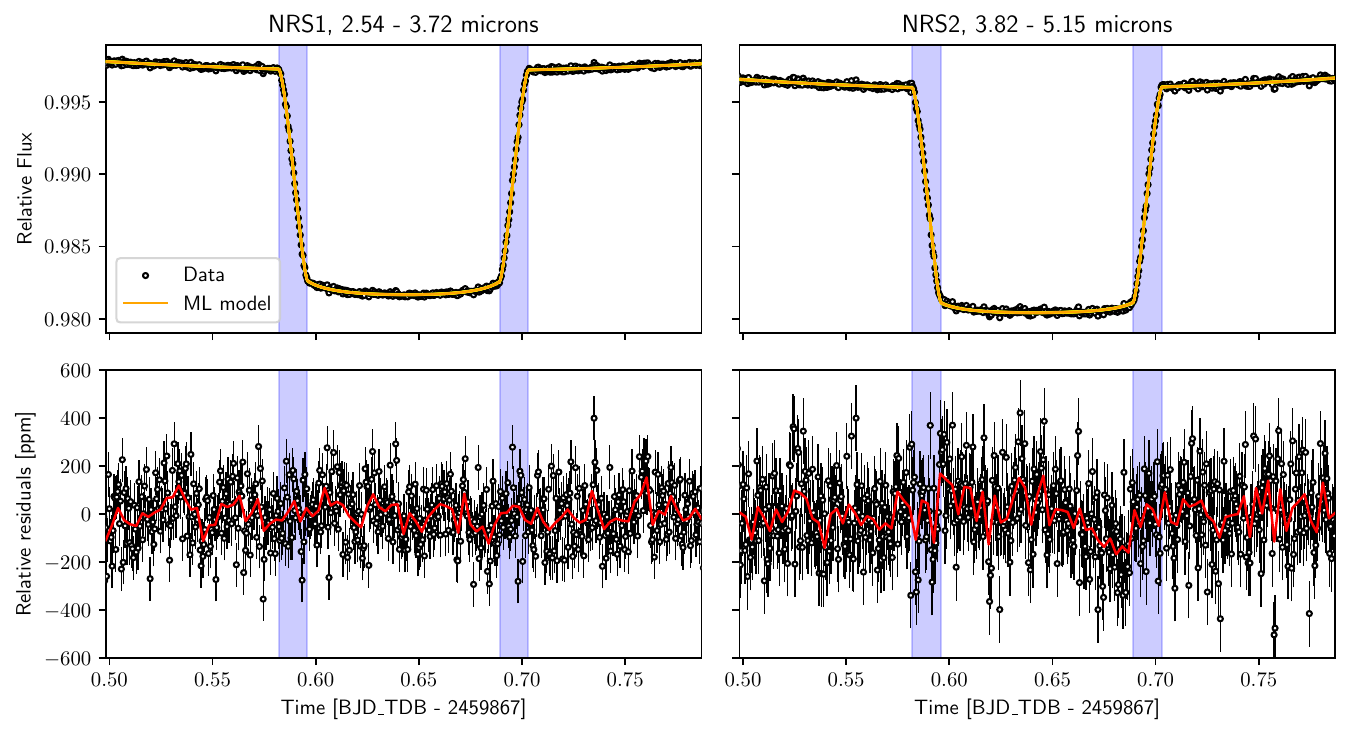}
	\caption{Results of the simultaneous MCMC on the white transit-only light curves of the NRS1 (\textit{left panels}) and NRS2 (\textit{right panels}) detectors. The shaded blue regions indicate the times of ingress and egress calculated according to \cite{winn10}. \textit{Top panels:} the black circles show the data and the orange lines show the model light curves applying the ML set of parameters. The faint yellow lines show models calculated using 100 random draws from the final parameter samples. \textit{Bottom panels:} the relative residuals between the data and the ML model are shown as black circles with the error bars of the data propagated onto the relative residuals. The residuals binned to a lower resolution in time are plotted using red solid lines.}
	\label{fig:white_lightcurves}
\end{figure*}

\begin{figure*}[htbp!]
    \centering
	\includegraphics[width=\hsize]{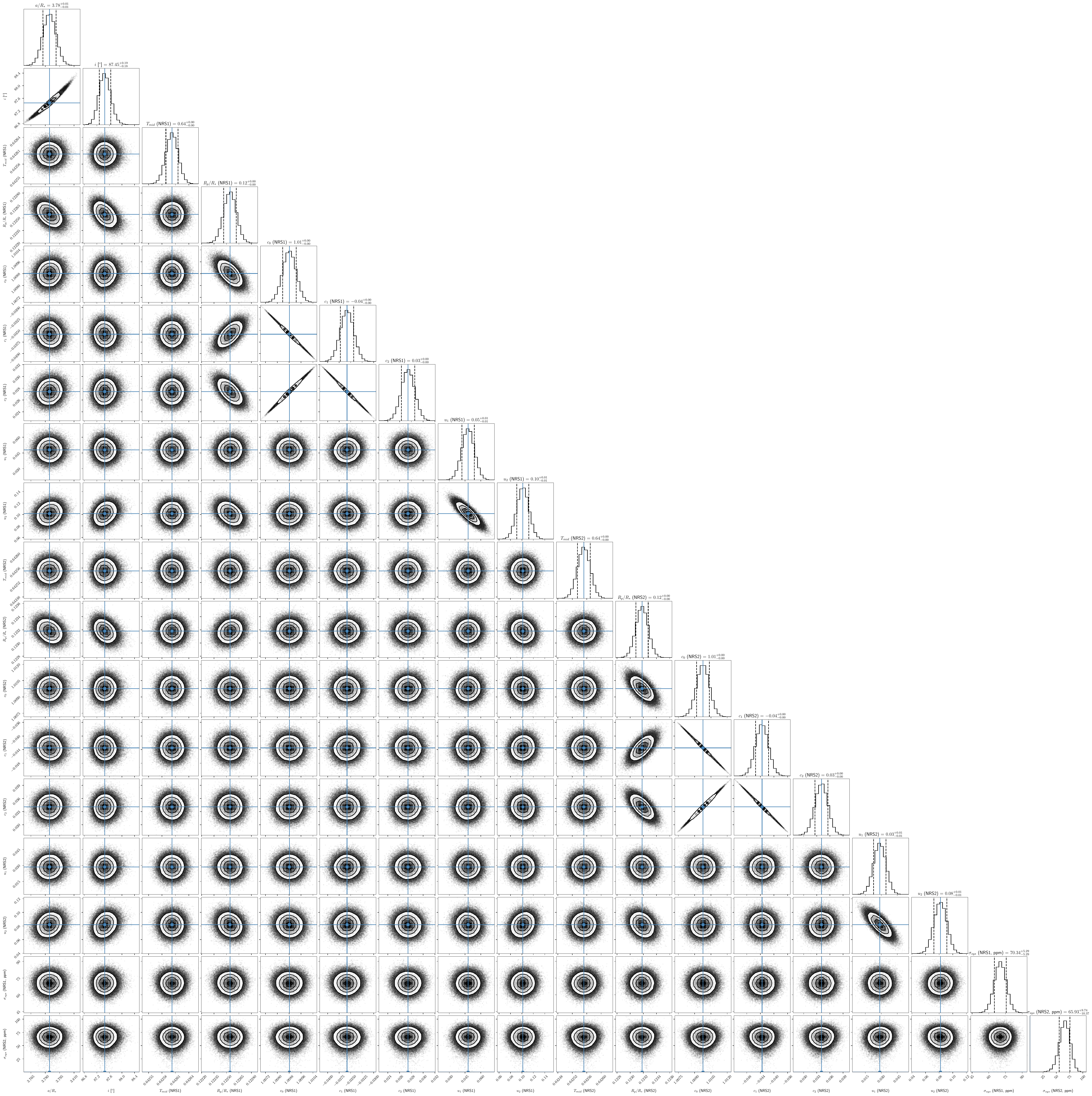}
	\caption{Posterior probabilities for the fit parameters of the simultaneous fit to both detectors' white light curves. For a description of each parameter, see Table \ref{tab:white_fit}.}
	\label{fig:white_corner}
\end{figure*}

To refine orbital parameters that are independent of the observed wavelengths but important for transit observations, we wavelength-integrated the light curves over the NRS1 and NRS2 detectors separately to create two white light curves we then analyzed jointly. We fit the two white light curves simultaneously using the transit model (Equation \ref{eq:transit_model}), employing the same orbital inclination and semi-major axis for both light curves and varying all other transit model and baseline parameters between the two light curves. We first conducted a preliminary least-squares fit and then performed sigma clipping, removing all data points that deviated from the preliminary model by more than $5\sigma$. This sigma clipping only resulted in the removal of one out of 650 data points of the NRS1 light curve and no removed data point of the NRS2 light curve. Then, we ran another least-squares fit and used the Markov-chain Monte Carlo (MCMC) package \texttt{emcee} \citep{emcee} to sample the parameters' posterior distributions, including a systematic white noise term ($\sigma_{sys}$), which was added to the Poisson data uncertainties derived from the reduction pipeline in quadrature. For the MCMC, we set wide uniform priors on all model parameters except for the quadratic limb darkening coefficients, for which we adopted normal priors from the \texttt{ExoTIC-LD} package \citep{exoticld} using the three-dimensional Stagger stellar grid \citep{staggergrid}. The resulting white light curve data and maximum likelihood (ML) models are shown in Figure \ref{fig:white_lightcurves}, the parameters' priors and posterior distributions are summarized in Table \ref{tab:white_fit} and the full parameters' posteriors are shown in Figure \ref{fig:white_corner}.
\begin{table*}[htbp!]
	\caption{Inputs and results for the light-curve model parameters for fitting the white light curves using MCMC.}
	\label{tab:white_fit}
	\centering
	\begin{tabular}{l c c c r}
		\hline\hline
		\textbf{Parameter} & \textbf{Description} & \textbf{Detector} & \textbf{Prior} & \textbf{Posterior} \\ \hline
		$a/R_*$ & Ratio of the semi-major axis to the star's radius & Joint & $\mathcal{U}(0,\infty)$ & $3.7844^{+0.0068}_{-0.0068}$ \\
		$i$ & Orbital inclination & Joint & $\mathcal{U}(0^\circ,90^\circ)$ & $\left(87.45^{+0.19}_{-0.18}\right)^\circ$ \\
		$T_\text{mid}$ & Transit midtime [BJD\_TDB$ - 2459867$] & NRS1 & $\mathcal{U}(-\infty,\infty)$ & $0.642602^{+0.000014}_{-0.000014}$ \\
		$T_\text{mid}$ & Transit midtime [BJD\_TDB$ - 2459867$] & NRS2 & $\mathcal{U}(-\infty,\infty)$ & $0.642552^{+0.000018}_{-0.000018}$ \\
		$R_p/R_*$ & Ratio of the planet's to the star's radius & NRS1 & $\mathcal{U}(-\infty,\infty)$ & $0.122542^{+0.000077}_{-0.000077}$ \\
		$R_p/R_*$ & Ratio of the planet's to the star's radius & NRS2 & $\mathcal{U}(-\infty,\infty)$ & $0.123180^{+0.000093}_{-0.000093}$ \\
		$u_1$ & Linear limb darkening coefficient & NRS1 & $\mathcal{N}(0.066, 0.012)$ & $0.0479^{+0.0059}_{-0.0060}$ \\
		$u_1$ & Linear limb darkening coefficient & NRS2 & $\mathcal{N}(0.056, 0.009)$ & $0.0286^{+0.0058}_{-0.0057}$ \\
		$u_2$ & Quadratic limb darkening coefficient & NRS1 & $\mathcal{N}(0.11, 0.02)$ & $0.102^{+0.010}_{-0.010}$ \\
		$u_2$ & Quadratic limb darkening coefficient & NRS2 & $\mathcal{N}(0.092, 0.012)$ & $0.0819^{+0.0094}_{-0.0095}$ \\
		$c_0$ & Constant baseline coefficient (see Equation \ref{eq:transit_model}) & NRS1 & $\mathcal{U}(-\infty,\infty)$ & $1.00881^{+0.00046}_{-0.00045}$ \\
		$c_0$ & Constant baseline coefficient (see Equation \ref{eq:transit_model}) & NRS2 & $\mathcal{U}(-\infty,\infty)$ & $1.00977^{+0.00059}_{-0.00059}$ \\
		$c_1$ & Linear baseline coefficient (see Equation \ref{eq:transit_model}) & NRS1 & $\mathcal{U}(-\infty,\infty)$ & $-0.0358^{+0.0014}_{-0.0015}$ \\
		$c_1$ & Linear baseline coefficient (see Equation \ref{eq:transit_model}) & NRS2 & $\mathcal{U}(-\infty,\infty)$ & $-0.0435^{+0.0019}_{-0.0019}$ \\
		$c_2$ & Quadratic baseline coefficient (see Equation \ref{eq:transit_model}) & NRS1 & $\mathcal{U}(-\infty,\infty)$ & $0.0274^{+0.0011}_{-0.0011}$ \\
		$c_2$ & Quadratic baseline coefficient (see Equation \ref{eq:transit_model}) & NRS2 & $\mathcal{U}(-\infty,\infty)$ & $0.0341^{+0.0015}_{-0.0015}$ \\
		$\sigma_{sys}$ & Systematic noise term & NRS1 & $\mathcal{U}(0,\infty)$ & $\left(70.3^{+5.3}_{-5.3}\right)$\,ppm \\
		$\sigma_{sys}$ & Systematic noise term & NRS2 & $\mathcal{U}(0,\infty)$ & $\left(65.9^{+9.7}_{-10.4}\right)$\,ppm \\ \hline
	\end{tabular}
    \tablecomments{Uncertainties reported here are $1\sigma$. In the prior column, $\mathcal{N}$ stands for a Gaussian with $\mu$ and $\sigma$ given in parentheses and $\mathcal{U}$ is a uniform prior with the lower and upper edges given in parentheses.}
\end{table*}

\begin{figure*}[htbp!]
    \centering
	\includegraphics[width=\hsize]{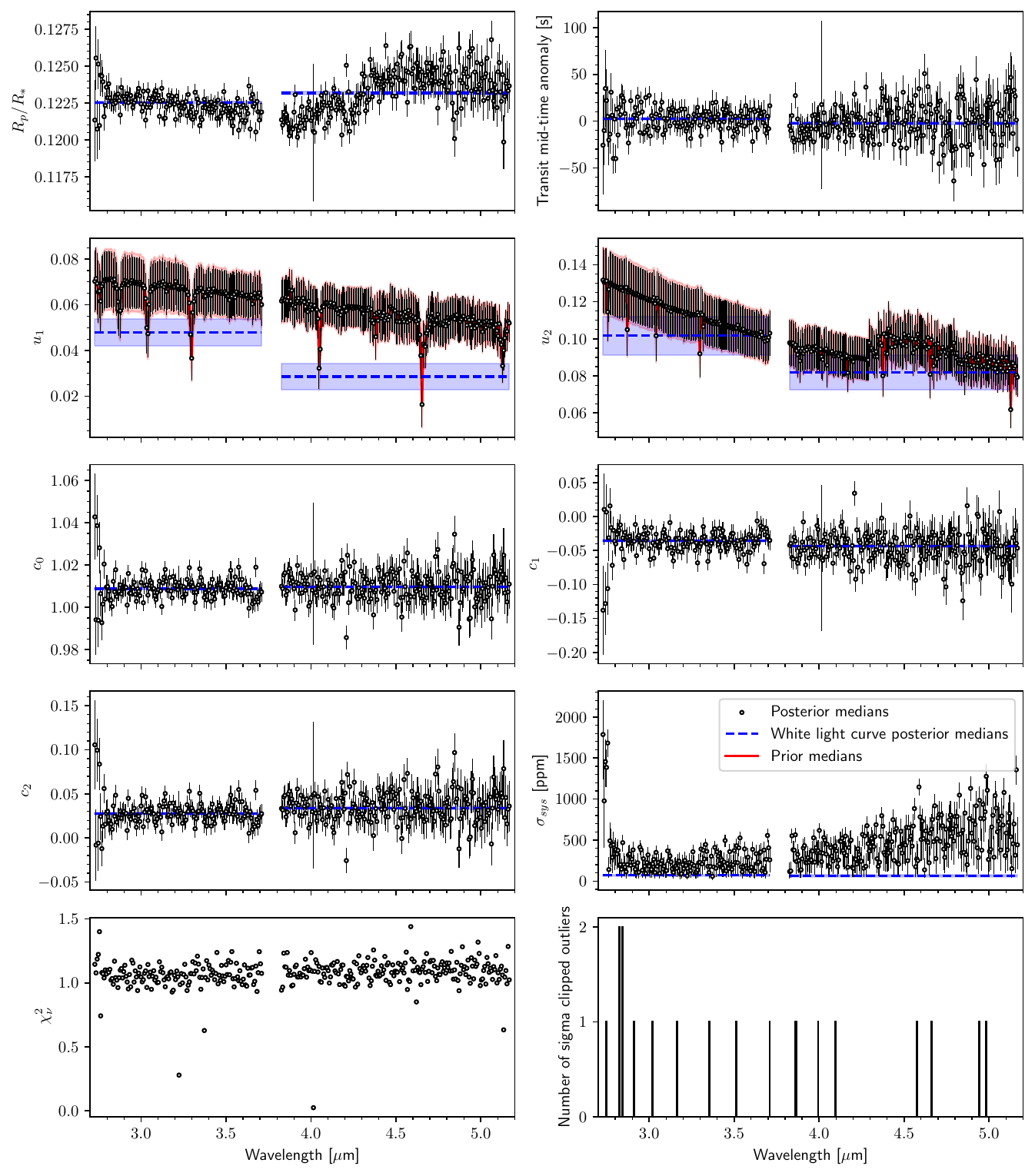}
	\caption{MCMC results for the fits to the spectrophotometric transit-only light curves at $R\sim 600$. For the fit parameters, we indicate the medians and $1\sigma$ intervals of the white light curve fit's posteriors (see Table \ref{tab:white_fit}) for both the NRS1 and NRS2 detectors using dashed blue lines and shaded blue regions. For the limb darkening parameters ($u_1$ and $u_2$), we adopted Gaussian priors, whose $\mu$ and $\sigma$ are shown using solid red lines and red shadings, respectively. The transit midtimes were subtracted by their median.}
	\label{fig:spectroscopic_fit_parameters}
\end{figure*}

\begin{figure}[htbp!]
    \centering
	\includegraphics[width=\hsize]{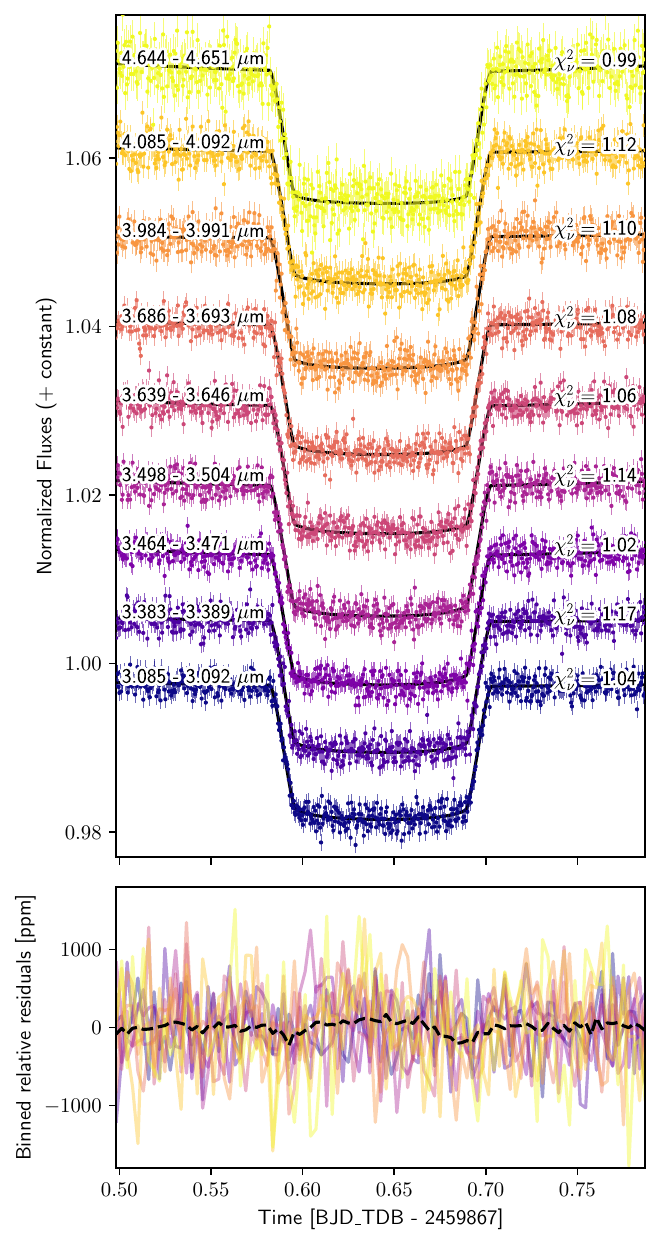}
	\caption{Selection of the spectrophotometric transit-only light curves at a spectral resolution of $R\sim 600$ and respective models. The selected spectrophotometric light curves are the 10th, 20th, 30th, 40th, 50th, 60th, 70th, 80th, and 90th percentile light curves in $\chi_\nu^2$. \textit{Upper panel:} colored dots with error bars show the data and solid black lines depict the model light curves generated employing the median values of the parameters' samples explored using MCMC. The wavelength ranges over which the light curves were integrated are given on the left and the $\chi_\nu^2$ of each model fit is given on the right. \textit{Lower panel:} the relative residuals between the selected data and the models binned to a lower resolution in time are shown using solid lines in the color that corresponds to the light curve in the upper panel they belong to. The dashed black line shows the median of the binned residuals of all spectrophotometric light curves (not just the ones plotted in the upper panel).}
	\label{fig:spectroscopic_lightcurves}
\end{figure}

\begin{figure}[htbp!]
    \centering
	\includegraphics[width=\hsize]{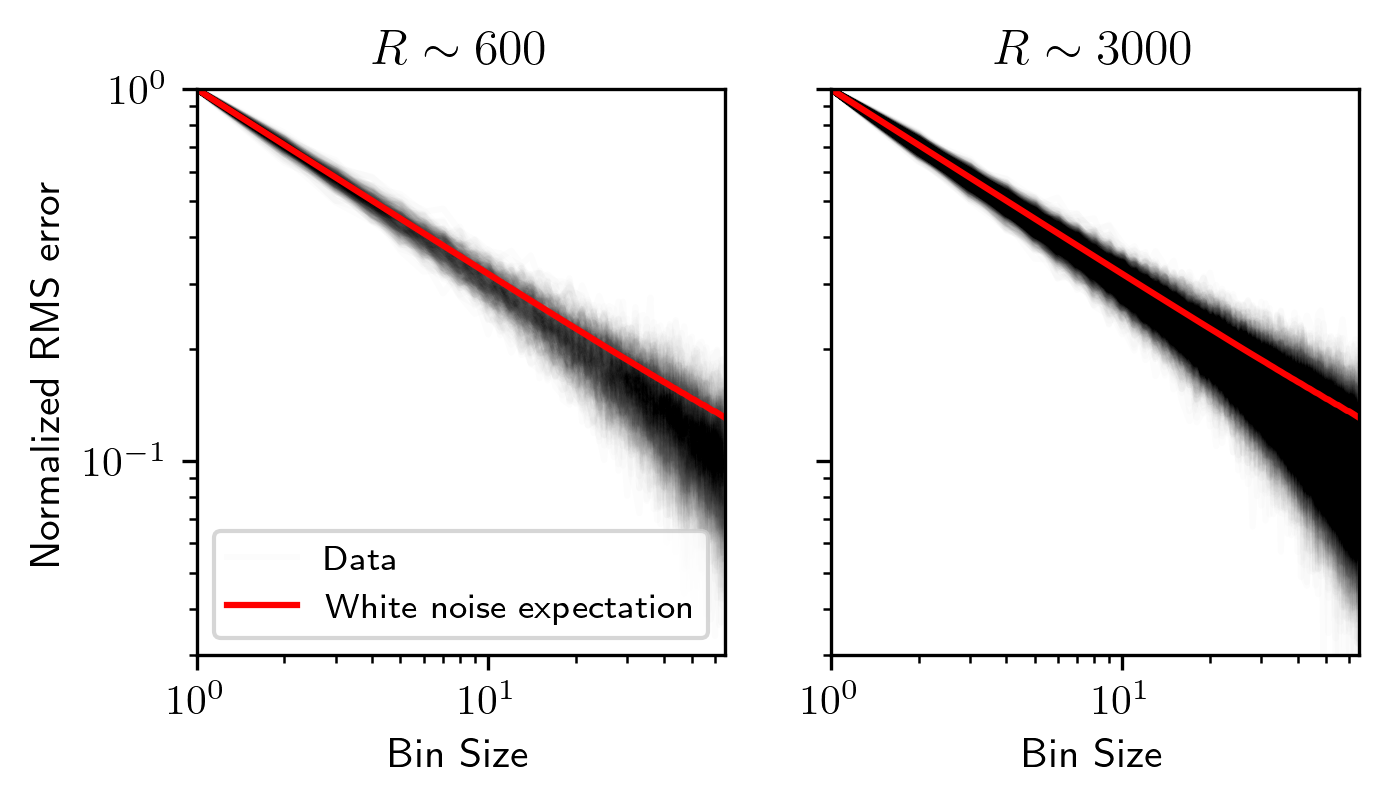}
	\caption{Allan deviation plots for the spectrophotometric transit-only light curves at both spectral resolutions. The transparent black lines show the root mean square (rms) residuals between the data and the models as a function of the number of bins and the red lines show the rms that would be expected if there were only white noise in the data.}
	\label{fig:allan_plot}
\end{figure}

After modeling the white light curves, we divided the time series of the one-dimensional spectra into finer wavelength bins to generate spectrophotometric light curves. We generated these light curves at two different wavelength binning levels: first, binning over 10 detector pixel columns per spectroscopic channel, leading to a spectral resolution of $R\sim 600$ and second, binning over two detector pixel columns, which results in a spectral resolution of $R\sim 3000$, both resolutions varying with wavelength. Finally, we fit the transit model (Equation \ref{eq:transit_model}) to each light curve individually, fixing the semi-major axis and orbital inclination to their ML values from the white light curve fit. Thus, the fit parameters for the spectrophotometric light curves were the baseline parameters ($c_0$, $c_1$ and $c_2$) as well as the planetary radius relative to the stellar radius ($R_p/R_*$), the transit midtime ($T_\text{mid}$) and the two limb darkening coefficients ($u_1$ and $u_2$). As in the white light curves' fit, we first fit the light curves using a least-squares approach, performed sigma clipping with a five sigma threshold and conducted another least-squares fit. The number of sigma-clipped data points for each spectrophotometric light curve is shown in Figure \ref{fig:spectroscopic_fit_parameters}. Then, we explored the parameter posteriors using MCMC for which we adopted normal priors calculated using the \texttt{ExoTIC-LD} package for the limb darkening coefficients and set wide uniform priors on all other parameters (as in the white light curves' fit; see Table \ref{tab:white_fit}). A selection of the spectrophotometric light curves at $R\sim 600$ and the respective light-curve models adopting the median values of all parameters are plotted in Figure \ref{fig:spectroscopic_lightcurves}. The Allan deviation plots for both the $R\sim 600$ and $R\sim 3000$ transit light curves are shown in Figure \ref{fig:allan_plot} and the posteriors for all fit parameters of all light curves are plotted in Figure \ref{fig:spectroscopic_fit_parameters}.

To examine the impact of fitting for the transit midtime in the spectrophotometric light curves, we also ran a separate fit of both the white and spectrophotometric light curves with a joint transit midtime in the white light curves' fit and fixed the transit midtimes of the spectrophotometric light curves to the white light curve fit's ML value. This only changed the medians of the transit depths by $<0.5\,\sigma$ and the size of the error bars by $\sim 0.1\,$ppm and thus, we analyze the transmission spectrum with free transit midtimes here.

\subsection{Comparison between transit-only and phase-curve analysis} \label{subsec:reduction-comparison}
\begin{figure*}[htbp!]
    \centering
	\includegraphics[width=\hsize]{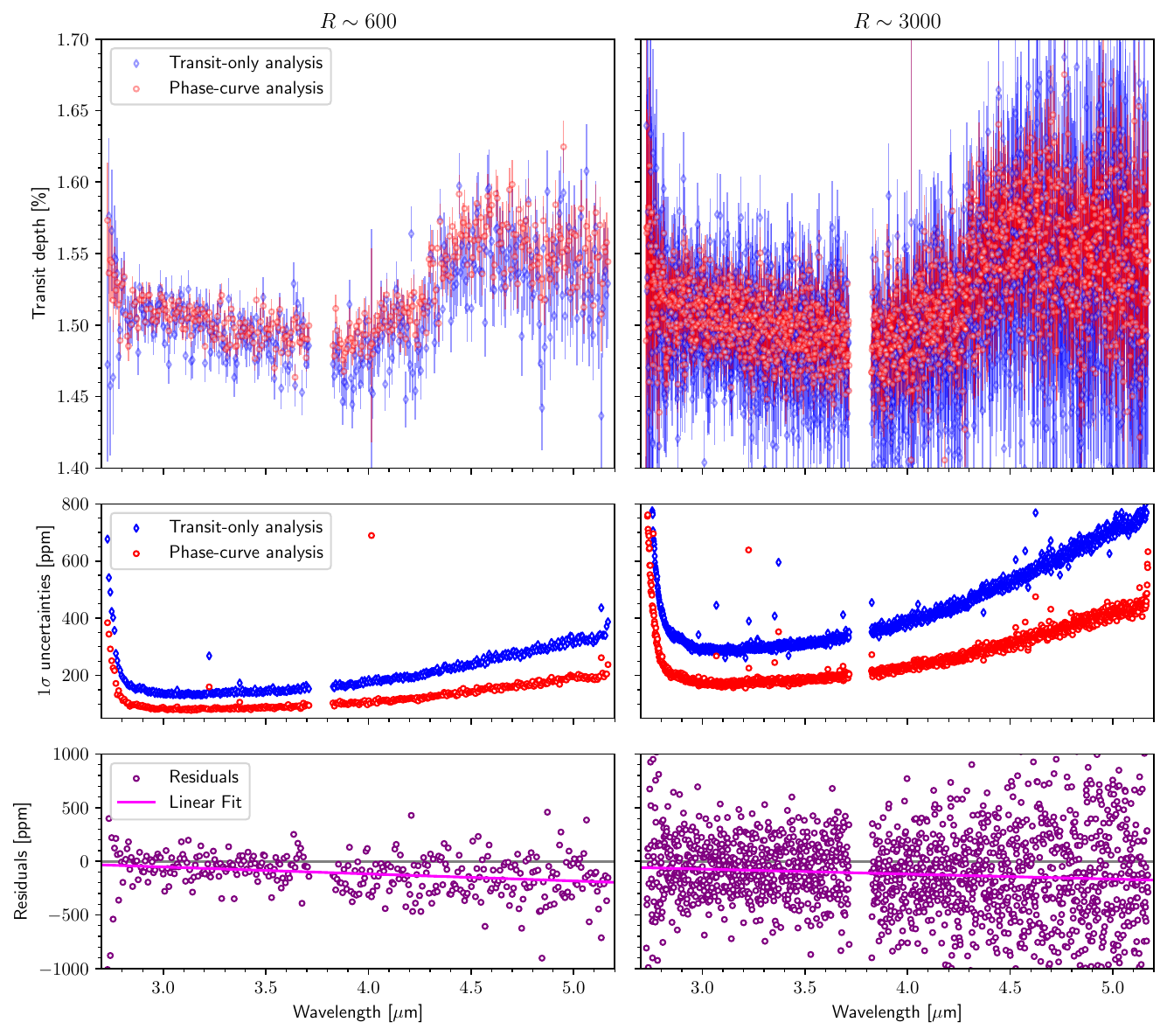}
	\caption{WASP-121\,b's transmission spectra from the transit-only analysis and the phase-curve analysis \citep{EvansSoma25} at two different spectral resolutions. \textit{Upper panels:} the blue diamonds and red circles show the transit-only and phase-curve transmission spectra, respectively. \textit{Middle panels:} the blue diamonds and red circles show the $1\sigma$ uncertainties of the transit-only and phase-curve transmission spectra, respectively. \textit{Lower panels:} the purple circles show the transit depths inferred from the transit-only analysis subtracted by the transit depths from the phase-curve analysis. Horizontal gray lines at zero were added to guide the eye. The solid magenta lines present linear fits ($f(\lambda)=c_0+c_1\lambda$) to the plotted residuals. For $R\sim 600$, $c_0 \approx 142$\,ppm and $c_1 \approx -65$\,ppm$\,\mu$m$^{-1}$ and for $R \sim 3000$, $c_0 \approx 68$\,ppm and $c_1 \approx -47$\,ppm$\,\mu$m$^{-1}$. The spectra displayed in this figure are available for download from the online journal.}
	\label{fig:transmission_spectra}
\end{figure*}

WASP-121\,b's transmission spectra derived from our transit-only analysis and presented as part of the phase-curve analysis \citep{EvansSoma25} are depicted in Figure \ref{fig:transmission_spectra} along with the residuals between them. The overall shape of both spectra is very similar; however, there are two notable differences between them: First, the transit-only analysis generally returns smaller transit depths and this offset increases with wavelength (see Figure \ref{fig:transmission_spectra}, lowest panels). And second, the uncertainties of the transit-only transmission spectrum are about $70\,\%$ larger than the ones of the phase-curve transmission spectrum (see Figure \ref{fig:transmission_spectra}, middle panels).

In the transit-only analysis, the increasing underestimation of the transit depth with wavelength as compared to the phase-curve analysis (see Figure \ref{fig:transmission_spectra}, lower panels) most likely originates in the contamination of the transit light curves with the planet's nightside emission. In the phase-curve analysis, the planet's emission as a function of phase is accounted for while the transit-only analysis has no leverage into separating stellar and planetary fluxes, leading to a bias toward lower transit depths that gets stronger with an increasing planetary-to-stellar flux ratio (see, e.g., \citealp{kipping10}). As the planetary-to-stellar flux ratio generally increases with wavelength, this effect leads to greater underestimation of the transit depth with wavelength. To estimate the amplitude of the transit depth underestimation of our transit-only analysis, we fit first-order polynomials to the residuals between both reduction pipelines' transmission spectra. For $R\sim 600$ ($R\sim 3000$), the fit line goes from about $-35$\,ppm ($-47$\,ppm) at $\lambda = 2.73\,\mu$m to $\approx -193$\,pmm ($-174$\,ppm) at $\lambda = 5.17\,\mu$m. For WASP-121\,b's transmission spectrum observed with JWST/NIRSpec, \cite{morello21} estimated the bias introduced by nightside emission to be about $(266\pm 127)$\,ppm, which depends on the adopted model parameters such as the planet's nightside temperature. As the difference in transit depths between our two data reductions increases with wavelength and comes out at a similar magnitude as estimated by \cite{morello21}, the bias introduced by the planet's nightside emission is likely the dominant source of the observed wavelength-dependent offset.

For both resolutions $R\sim 600$ and $R\sim 3000$, the uncertainties of the transit-only transmission spectrum are $\sim 70\,\%$ larger than the phase-curve transmission spectrum's uncertainties (see Figure \ref{fig:transmission_spectra}, middle panels). To examine the origin of the size of the error bars, we ran the transit-only analysis with one fitting parameter (either $T_\text{mid}$, $u_1$, $u_2$, $c_2$ or $\sigma_{sys}$; see Table \ref{tab:white_fit}) fixed at a time and inspected the uncertainties on the transit depth. The only significant decrease of the transit depth uncertainties we observed during that exercise was when $c_2$ was fixed to zero and thus, when a linear instead of a quadratic baseline was adopted. From this observation, we suspect that the increase of transit depth uncertainties of the transit-only analysis as compared to the phase-curve analysis mainly originates in the correlation between the quadratic baseline and the transit depth. In the posteriors of the white light curve fit (Figure \ref{fig:white_corner}), there is a clear negative correlation between $c_2$ and $R_p/R_*$, meaning that a curvier baseline leads to a smaller planetary radius as a curvier baseline reaches lower fluxes during the transit. This correlation appears to be strong enough to overwhelm the correlation between $c_0$ and $R_p/R_*$, which is negative in our posteriors (see Figure \ref{fig:white_corner}), but is positive when a linear baseline is used. Hence, our hypothesis is that the transit-only analysis with its narrower phase coverage of WASP-121\,b's phase-curve delivers weaker constraints on the phase-curve shape which subsequently propagate onto the transit depth uncertainties via the degeneracy with $c_2$. We refrained from neglecting the out-of-transit curvature of the observations by adopting a linear instead of a quadratic baseline into our model (Equation \ref{eq:transit_model}), since the light curves cannot be adequately fit with a linear baseline due to the strong curvature (see Figures \ref{fig:white_lightcurves} and \ref{fig:spectroscopic_lightcurves}). In the test run using a linear baseline in the light curve model, we observed a systematic overestimation of the transit depth as compared to the run with a quadratic baseline and an increase of the overestimation with wavelength. This bias is caused by the neglected phase-curve curvature that becomes stronger as the planetary-to-stellar flux ratio increases.

In the following analysis of WASP-121\,b's transmission spectrum, we applied atmospheric retrievals to both the transit-only and the phase-curve transmission spectrum to investigate the impact of the planet nightside emission's contamination of the light curves and the weaker constraints on the phase-curve in the transit-only analysis. Since the phase-curve analysis, however, uses all taken observations to constrain the planet's emission as a function of time and the correlated transit depth, we consider that approach preferable and draw conclusions primarily driven by the results of that data reduction.

\subsection{Limb asymmetries}
When interpreting exoplanet transmission spectra, it has traditionally been assumed that the terminator of the planet is homogeneous. However, with JWST-quality data it may be possible to discern terminator inhomogeneities driven by differences in temperature and chemical composition of the morning and evening terminators of the planets, as recently demonstrated for WASP-39\,b \citep{wasp39} and WASP-107\,b \citep{wasp107}. We wanted to investigate whether a uniform terminator model is a sufficient approach for the atmosphere of WASP-121\,b. We explored this by applying an asymmetric model to both the white and the spectroscopic light curves using the package \texttt{catwoman} \citep{catwoman,catwoman2} with the nested sampling algorithm \texttt{dynesty} \citep{dynesty}.

The white and spectrophotometric light curves of the transit were extracted from the phase-curve light curves \citep{EvansSoma25} that were analyzed to correct for planetary emission contamination. The \texttt{catwoman} package models the transiting planet as two stacked semicircles with different radii and an angle of rotation between the terminator and the direction of propagation of the planet. This angle was set to $90^\circ$ and a quadratic limb darkening law was assumed. The limb darkening coefficients were obtained from the \texttt{ExoTIC-LD} package \citep{exoticld} using the Stagger stellar grid \citep{staggergrid} and were fixed to those values.  We used a semi-major axis of $a/R_* = 3.7844$ and an inclination of $87.45^{\circ}$ as reported in Table \ref{tab:white_fit}. In the white light curves, we fit the model with a variable transit midtime. However, since no significant difference in transit midtime was observed between the \texttt{catwoman} and \texttt{batman} models, we chose to fix it for the spectroscopic light curves at 2459867.642577\,BJD\_TDB (see Table \ref{tab:white_fit}). We fit a model with two semicircle radii for the morning and evening of the planet, linear and offset terms for the baseline flux, and a systematic noise term. We fixed the integration step scale for the \texttt{catwoman} models to significantly reduce the fitting time. For the default maximum error that is produced by \texttt{catwoman} (1\,ppm), the optimal integration step scale is \texttt{fac = 0.031}. We set it to 0.01 in our fits, which reduces the maximum error in the model by a factor of approximately 10. A smaller integration step scale was explored and it led to a substantial increase in the fitting time, with a negligible decrease in the error bars. We used a total of 1200 live points for each of the model fits.

\begin{figure*}[htbp!]
    \centering
	\includegraphics[width=\hsize]{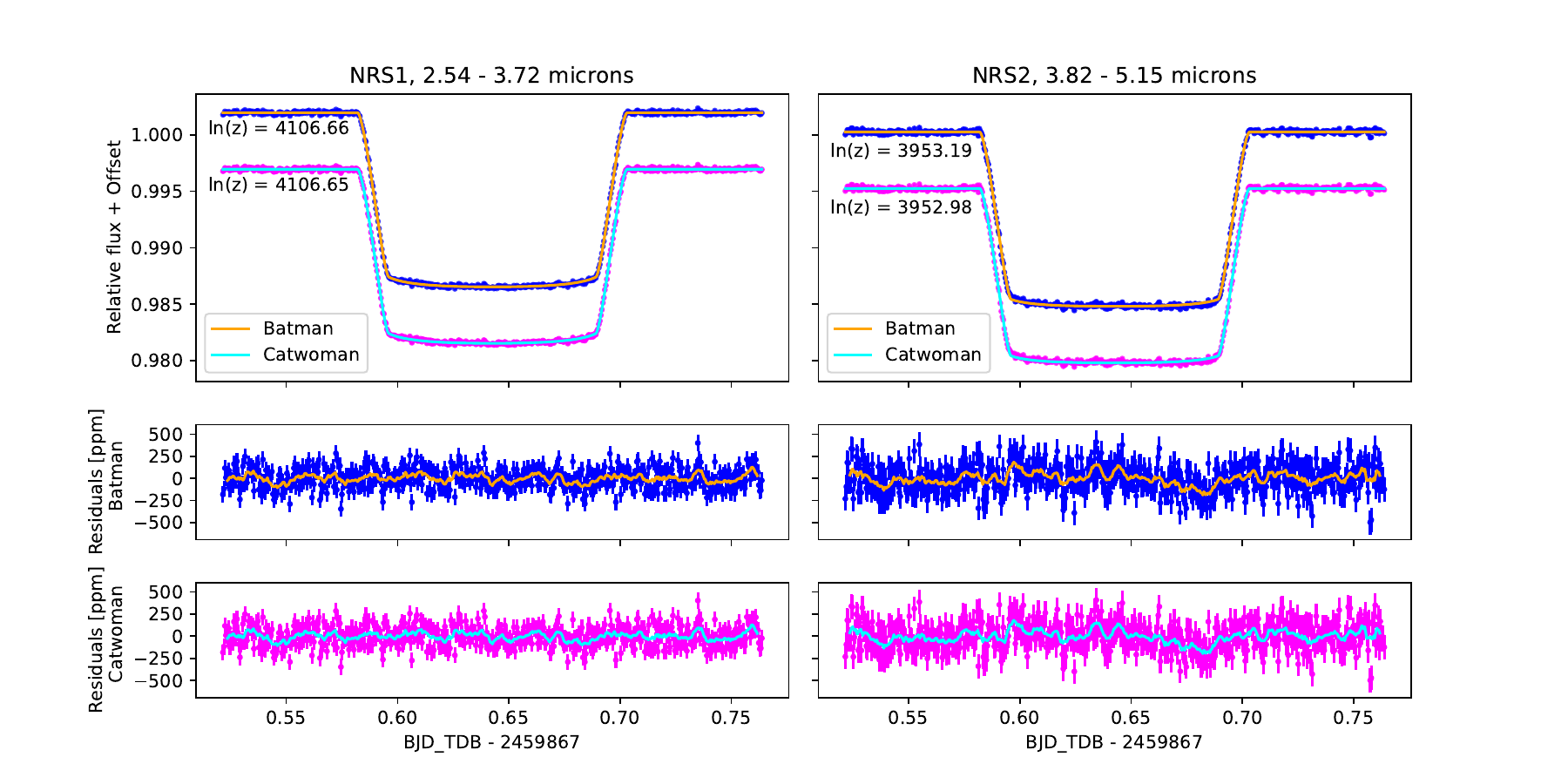}
	\caption{Fits to the white light curves using \texttt{batman} and \texttt{catwoman} models.}
	\label{fig:catwoman_white}
\end{figure*}
\begin{figure*}[htbp!]
    \centering
	\includegraphics[width=\hsize]{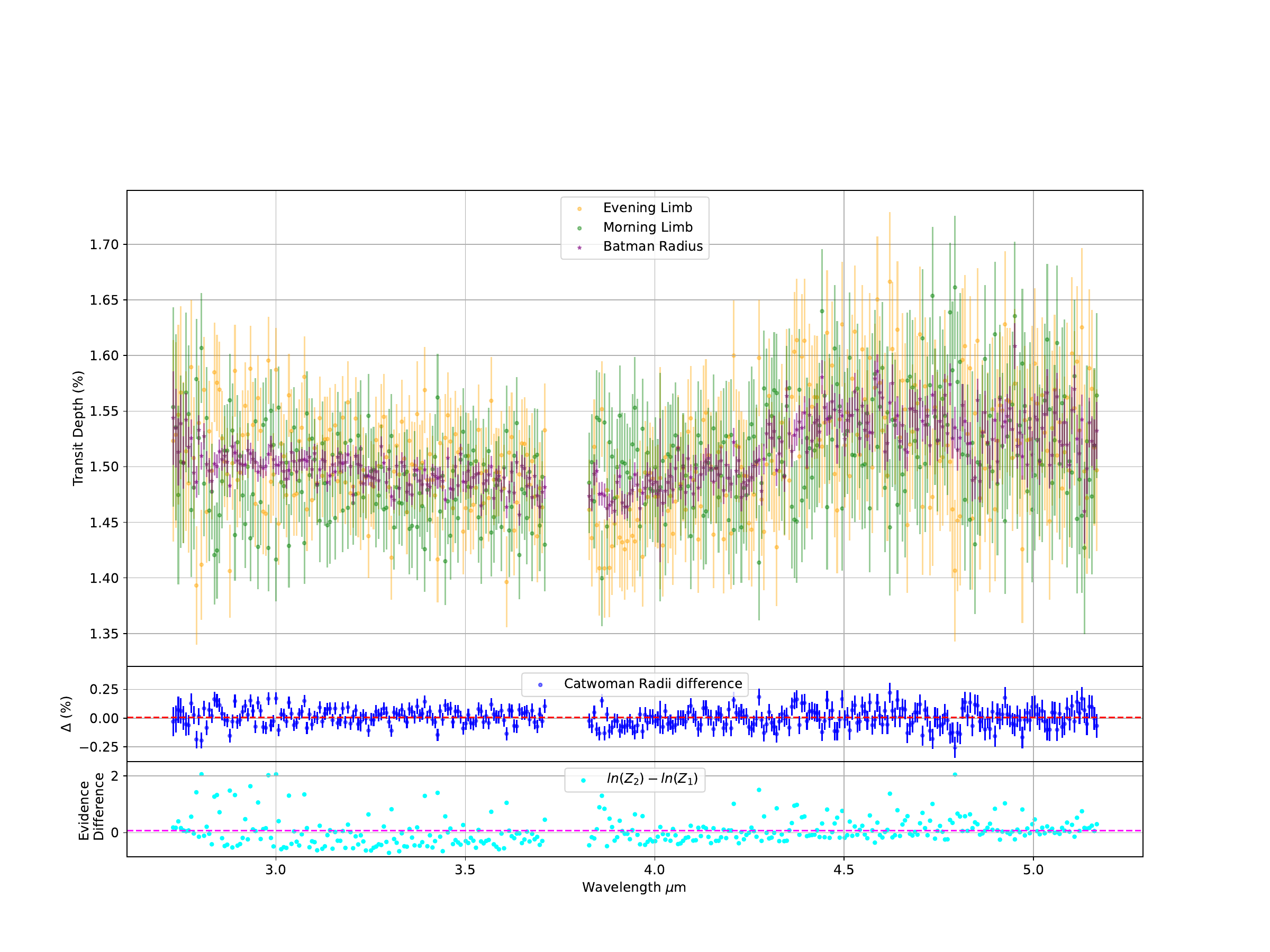}
	\caption{Comparison of the transmission spectra for the \texttt{batman} and \texttt{catwoman} models.\textit{Top panel:} trailing and leading limb depths compared to the depths found from the \texttt{batman} model fitting. \textit{Middle panel:} differences between the trailing and leading limb depths. The red line shows the mean value of $0.006 \%$. \textit{Bottom panel:} difference in logarithmic evidence between the \texttt{catwoman} model ($ \ln\mathcal(Z_2)$) and the \texttt{batman} model ($ \ln\mathcal(Z_1)$) as estimated from \texttt{dynesty}. The pink line shows the mean value of $0.06$.}
	\label{fig:catwoman}
\end{figure*}
The Bayesian evidence of the models for the white light curves of the NRS1 and NRS2 detectors (see Figure \ref{fig:catwoman_white}) is shown in Table \ref{tab:model_comparison} and demonstrates that the \texttt{batman} model is slightly favored in both of the detectors as the Bayes factors are less than one.  A comparison between the transmission spectra received from the \texttt{batman} and \texttt{catwoman} models is presented in Figure \ref{fig:catwoman}. Note that the semicircle radii have significantly larger uncertainties due to a strong negative correlation in the posterior (see also, e.g., \citealp{wasp39}). Additional tests were performed by setting the transit midtime as a free parameter. However, this increased the uncertainties in radius difference even further without any implications on the model comparison. The mean difference between the two radii is $0.006 \%$, showing no evidence of a larger evening atmosphere. Using nested sampling allows us to estimate the Bayesian evidence and compare the two models. For 195 out of 349 spectroscopic light curves, \texttt{batman} has higher evidence and for the rest, \texttt{catwoman} is preferred, with a maximum logarithmic evidence difference of 2.05, corresponding to a Bayes factor of 7.77, which is not significant enough for us to choose it as the more adequate model. Based on these results, we prefer a symmetric model for the transit-only analysis as there is insufficient evidence for the alternative.
\begin{table}[htbp!]
\caption{Logarithmic Bayesian Evidence for the two models for the light curves from the NRS1 and NRS2 detectors. }
\label{tab:model_comparison}
\centering
\begin{tabular}{ c c c c}
\hline \hline
   & $\mathbf{ \ln \mathcal(Z_1)}$ & $ \mathbf{\ln \mathcal(Z_2)}$ & \textbf{Bayes Factor} \\
\hline
\textbf{NRS1} & 4106.66 & 4106.65 & 0.990 \\
\textbf{NRS2} & 3953.19 & 3952.98 & 0.811 \\
\hline
\end{tabular}
\tablecomments{$\ln \mathcal(Z_1)$ and $ \ln \mathcal(Z_2)$ correspond to the evidence from the \texttt{batman} and \texttt{catwoman} models, respectively.}
\end{table}

\section{Atmospheric models} \label{sec:retrievals}
To identify atmospheric absorbers that contribute to WASP-121\,b's transmission spectrum between 2.7 and 5.2\,$\mu$m, we analyzed both the phase-curve transmission spectrum and the transit-only transmission spectrum at a spectral resolution of $R\sim 600$ using three different retrieval suites. One of these, the \texttt{NEMESIS} framework, includes an atmospheric model that allows different temperatures and chemistry between the planet's highly irradiated dayside and the cooler nightside, while the others (\texttt{ATMO} and \texttt{PETRA}) do not allow differences between hemispheres. Additionally, we compared the phase-curve's transmission spectrum to model spectra derived from postprocessed output of a GCM presented by \cite{pluriel20} to explore the possible impact of WASP-121\,b's three-dimensional nature on the transmission spectrum.

\subsection{\texttt{NEMESIS} (Free Chemistry)} \label{subsec:nemesis}
\texttt{NEMESIS} is a radiative transfer and retrieval package originally developed for solar system atmospheres, and more recently applied extensively to exoplanets (see, e.g., \citealt{lewis20,MikalEvans22}). It uses a fast correlated-k framework \citep{lacis91} for calculating molecular and atomic opacities, and the nested sampling \citep{skilling04} algorithm \texttt{PyMultiNest} \citep{feroz2009,Feroz2019,Buchner2014} to sample the parameter space and converge on a solution. For WASP-121\,b's transmission spectrum, we retrieved the abundances of H$_2$O, CO$_2$, CO, SiH, SiO and H$^-$, for which we adopted the opacity functions from \cite{ExoMol_H2O}, \cite{ExoMol_CO2}, \cite{15LiGoRo.CO}, \cite{ExoMolSiH}, \cite{barton2013} and \cite{john1988}, respectively. We used $R=3000$ k tables generated according to \cite{chubb2021} which were then channel-averaged to the resolution of the data, except for the H$^-$ absorption which is effectively continuum absorption. Additionally, we included collision-induced absorption from H$_2$ and helium according to \cite{borysow89,borysow90}, \cite{borysowfm89,borysow97,Borysow2001} and \cite{borysow02}.

\begin{figure}[htb]
   \centering
    \includegraphics[width=\hsize]{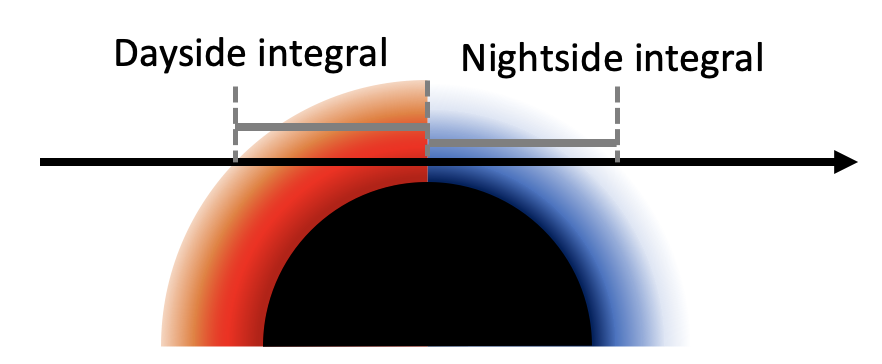}
    \caption{Schematic showing an example path for the \texttt{NEMESIS} dayside-and-nightside radiative transfer calculation. The dayside portion of the atmosphere is represented by shades of red and the nightside by shades of blue. The intensity of the color indicates the atmospheric density, and it can be seen that the density and therefore the pressure on the nightside drop off much more rapidly as a function of altitude. The model altitude coordinate is continuous across the day-night boundary, but the pressure coordinate is not.}
    \label{fig:nemesisschematic}
\end{figure}

To account for the expected thermochemical differences between WASP-121\,b's dayside and nightsides (see Section \ref{sec:introduction}), we implemented an atmospheric model that divides the planet into two hemispheres in the \texttt{NEMESIS} framework (see Figure \ref{fig:nemesisschematic}). We allowed the dayside and nightside components in this retrieval setup to have differences in their temperatures as well as their chemistry with the exception of CO, SiO and SiH which we assumed to be constant throughout the planet. For H$_2$O, H$_2$, CO$_2$ and H$^-$, we adopted separate dayside and nightside abundances. On the dayside, we additionally allowed H$_2$ and H$_2$O to thermally dissociate in the upper atmosphere by parameterizing their abundance profiles using
\begin{equation}
	X(P) =
	\begin{cases}
		X_{\mathrm{deep}} & \text{if } P\ge P_{\mathrm{knee}} \\
		X_{\mathrm{deep}} \left(P/P_{\mathrm{knee}}\right)^\alpha & \text{else}
	\end{cases}.
    \label{eq:knee_profile}
\end{equation}
Here, $P$ is pressure, $X_{\mathrm{deep}}$ is the molecule's abundance below the knee pressure level $P_{\mathrm{knee}}$, and $\alpha$ is a power-law index  describing the rate at which the abundance decreases with pressure above the knee pressure level. For H$_2$, we set $X_{\mathrm{deep}}=0.8547$, and assumed that all dissociated H$_2$ is converted to atomic hydrogen, so that the atomic hydrogen abundance increases as H$_2$ decreases. For H$_2$O and H$_2$ on the nightside and for all other molecules, we adopted pressure-independent mole fractions. Finally, we allowed for H$^-$ absorption on the dayside but not on the nightside, and we retrieved the volume mixing ratio of H$^-$. The free electron volume mixing ratio was fixed to be equal to that of H$^-$, as was the atomic hydrogen abundance in the deep atmosphere, where it is not augmented by dissociated H$_2$.

We modeled the dayside and nightside temperatures using a modified version of the `Guillot' profile \citep{guillot10}. The dayside pressure-temperature (P-T) profile was modeled using a reduced three-parameter version of the `Guillot' profile with a single optical stream; retrieved parameters are the infrared opacity (${\kappa}_{\mathrm{IR}}^{\mathrm{(day)}}$), the ratio of the optical opacity to the infrared opacity ($\gamma^{\mathrm{(day)}}={\kappa}_{\mathrm{vis}}^{\mathrm{(day)}}$/${\kappa}_{\mathrm{IR}}^{\mathrm{(day)}}$), and $\beta^{\mathrm{(day)}}$ which allows the effective irradiation temperature to be tuned. On the nightside, we made the assumption that $\beta^{\mathrm{(night)}}$ must be zero since no irradiation is received from the star; $\gamma^{\mathrm{(night)}}$ which governs the optical opacity, is also irrelevant due to the lack of short-wavelength radiation. We therefore described the nightside P-T profile using
\begin{equation}
    T(P) = \left(\frac{3}{4}\left(T_{\mathrm{int}}+a^{(1/4)}T_{\mathrm{day}}\right)^4 \left(\frac{2}{3} + \frac{\kappa_{\mathrm{IR}}^{\mathrm{(night)}} P}{g}\right)\right)^{(1/4)}, \label{eq:pt_night}
\end{equation}
where $T_{\mathrm{int}}$ is the internal temperature, set to 500 K, $T_{\mathrm{day}}$ is the temperature at the bottom of the model atmosphere on the dayside and the term $a^{(1/4)}T_{\mathrm{day}}$ represents the additional heat advected from the dayside in the deep atmosphere; $\kappa_{\mathrm{IR}}^{\mathrm{(night)}}$ is the infrared opacity on the nightside and $g$ is the gravitational acceleration. The two additional free parameters governing the nightside temperature are $a$, which tunes the amount of extra heat deposited in the deep atmosphere, and $\kappa_{\mathrm{IR}}^{\mathrm{(night)}}$. This parameterization does not allow for a thermal inversion on the nightside, however, we consider a noninverted temperature profile on the nightside a reasonable assumption for WASP-121\,b, as phase-curve observations using HST have delivered evidence for a decrease of the atmospheric temperature with height on the nightside \citep{MikalEvans22}.

\begin{figure*}[htbp!]
    \centering
	\includegraphics[width=\hsize]{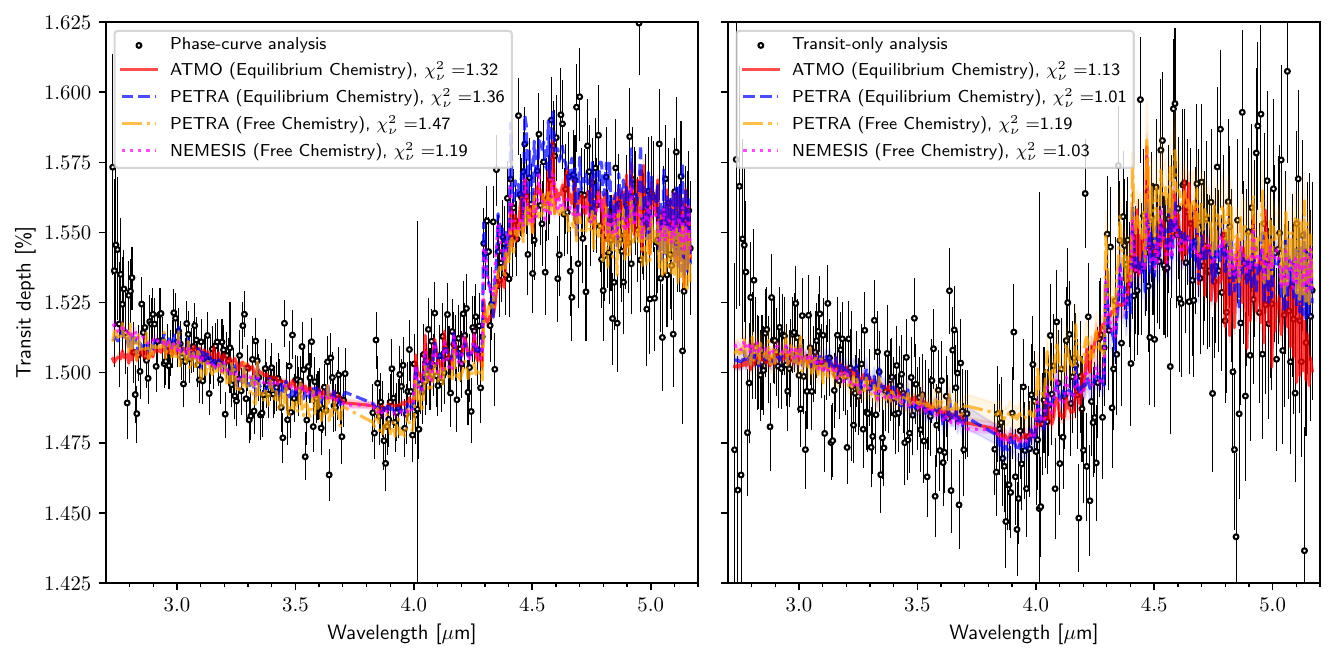}
	\caption{Phase-curve (\textit{left panel}) and transit-only (\textit{right panel}) transmission spectra at $R\sim 600$ along with the median model spectra of the applied retrievals. The $1\sigma$ intervals of the model spectra are illustrated using shaded regions.}
	\label{fig:retrieval-spectra}
\end{figure*}

\begin{figure*}[htbp!]
    \centering
	\includegraphics[width=\hsize]{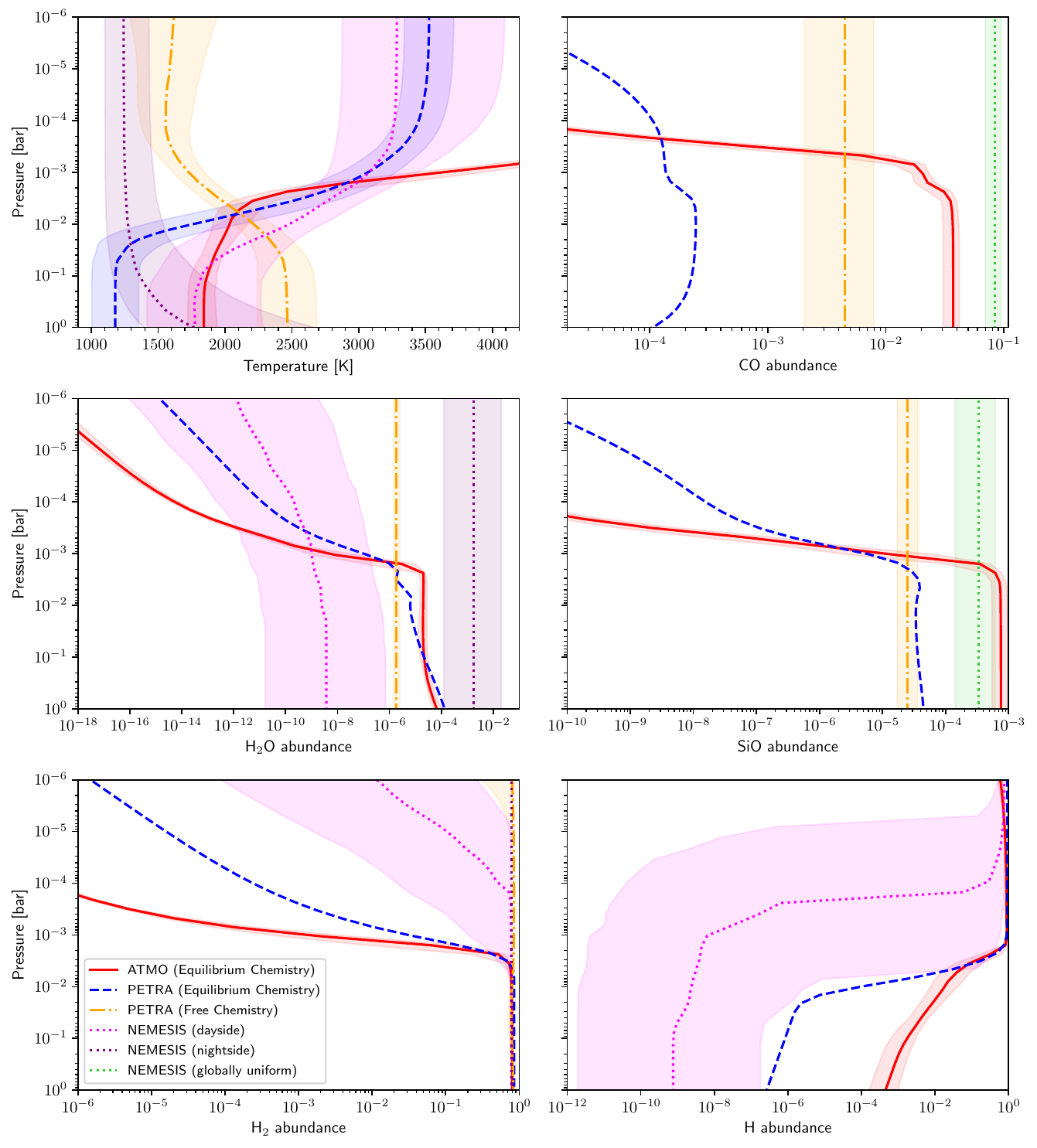}
	\caption{Results from the retrieval frameworks applied to WASP-121\,b's transmission spectrum obtained from the phase-curve analysis. The temperature and abundance profiles inferred from the different retrievals are depicted using lines with shaded regions indicating the extent of the corresponding $1\sigma$ regions. All \texttt{NEMESIS} results shown here were inferred from a simultaneous retrieval of both hemispheres. Any atmospheric variable varied between the hemispheres in that framework is presented using dotted magenta and purple lines, and the molecular abundances that are held identical in both hemispheres are shown using dotted green lines.}
	\label{fig:retrieval-atmospheres_phase-curve}
\end{figure*}

\begin{figure*}[htbp!]
    \centering
	\includegraphics[width=\hsize]{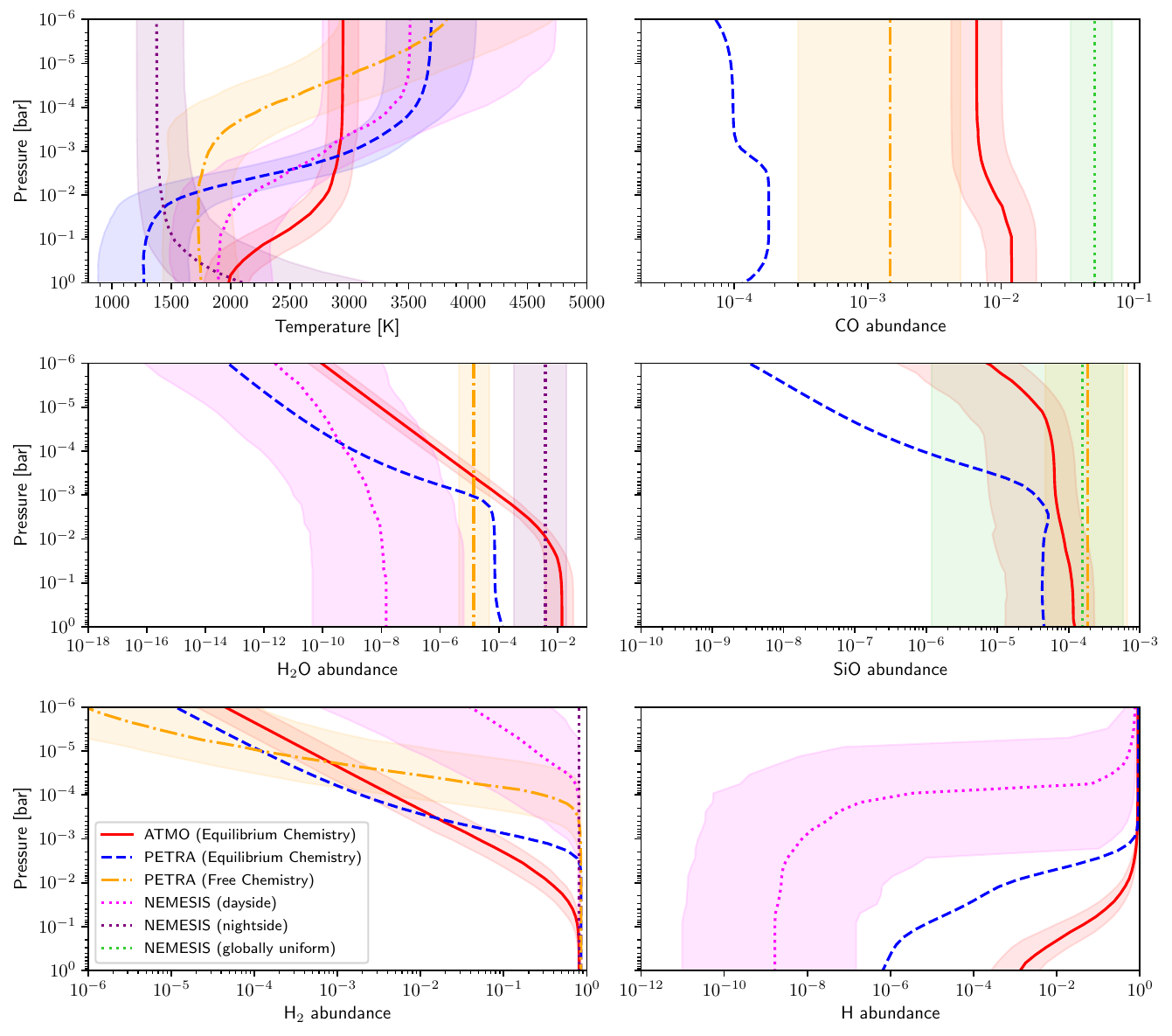}
	\caption{Same as Figure \ref{fig:retrieval-atmospheres_phase-curve}, but for the retrievals applied to the transit-only transmission spectrum.}
	\label{fig:retrieval-atmospheres_transit-only}
\end{figure*}

Table \ref{tab:nemesis} lists the priors we chose for all retrieval parameters as well as the posteriors we found in the retrievals on both data reductions' transmission spectra. For all fit parameters, both retrievals agree remarkably well. The fits to the data are very good for both the retrieval on the phase-curve transmission spectrum and the retrieval on the transit-only transmission spectrum, with $\chi_\nu^2=1.19$ and $\chi_\nu^2=1.04$, respectively (see Figure \ref{fig:retrieval-spectra}). In both retrievals, we recovered an inverted P-T profile for WASP-121\,b's dayside which increases from $T\sim 2000\,$K at $p\sim100\,$mbar to $T\sim3400\,$K at $p\sim0.1\,$mbar (see Figure \ref{fig:retrieval-atmospheres_phase-curve} for the retrieval on the phase-curve transmission spectrum and Figure \ref{fig:retrieval-atmospheres_transit-only} for the retrieval on the transit-only transmission spectrum). The nightside profile decreases from $T\sim2000\,$K in the deep atmosphere to $T\sim1350\,$K at $p\sim10\,$mbar, and is almost isothermal above this. The retrieved median H$_2$O abundances are about 6 orders of magnitude lower on the dayside than on the nightside.
\begin{table*}[htbp!]
	\caption{Inputs and results for the \texttt{NEMESIS} retrieval employing free chemical abundances, split by hemisphere.}
	\label{tab:nemesis}
	\centering
	\begin{tabular}{l c c r r}
		\hline\hline
        & & & \multicolumn{2}{c}{\textbf{Posteriors}} \\
		\textbf{Parameter} & \textbf{Hemisphere} & \textbf{Prior} & \multicolumn{1}{c}{\textbf{PC}} & \multicolumn{1}{c}{\textbf{TR}} \\ \hline
		WASP-121\,b's radius relative to Jupiter's radius & Both & $\mathcal{U}(1.2,2.0)$ & $1.69^{+0.01}_{-0.01}$ & $1.66^{+0.01}_{-0.01}$ \\
		Cloud top pressure in $\log_{10}(\text{bar})$ & Both & $\mathcal{U}(-6.0,1.0)$ & $-2.62^{+2.16}_{-2.12}$ & $-2.77^{+2.41}_{-1.82}$ \\
		Cloud opacity scaling in $\log_{10}$ & Both & $\mathcal{U}(-10.0,10.0)$ & $-4.66^{+3.44}_{-3.22}$ & $-4.30^{+2.59}_{-3.02}$ \\
		CO mole fraction in $\log_{10}$ & Both & $\mathcal{U}(-12.0,-1.0)$ & $-1.08^{+0.05}_{-0.08}$ & $-1.27^{+0.11}_{-0.13}$ \\
		SiH mole fraction in $\log_{10}$ & Both & $\mathcal{U}(-12.0,-1.0)$ & $-8.99^{+1.99}_{-1.82}$ & $-8.31^{+1.94}_{-2.35}$ \\
		SiO mole fraction in $\log_{10}$ & Both & $\mathcal{U}(-12.0,-1.0)$ & $-3.48^{+0.27}_{-0.37}$ & $-3.51^{+0.39}_{-0.47}$ \\
		$a$ (see Equation \ref{eq:pt_night}) & Both & $\mathcal{U}(0.0,0.15)$ & $0.08^{+0.04}_{-0.04}$ & $0.09^{+0.04}_{-0.04}$ \\
		$\beta^{(\mathrm{day})}$ (see \citealp{Line2012ApJ...749...93L,line13}) & Dayside & $\mathcal{U}(0,2.0)$ & $0.85^{+0.19}_{-0.18}$ & $0.88^{+0.19}_{-0.17}$ \\
		$\gamma^{(\mathrm{day})}$ in $\log_{10}$ (see \citealp{guillot10,Line2012ApJ...749...93L,line13}) & Dayside & $\mathcal{U}(-4.0,3.0)$ & $1.18^{+0.60}_{-0.56}$ & $1.14^{+0.75}_{-0.65}$ \\
		$\kappa_{\mathrm{IR}}^{(\mathrm{day})}$ in $\log_{10}$ (see \citealp{guillot10,Line2012ApJ...749...93L,line13}) & Dayside & $\mathcal{U}(-4.0,3.0)$ & $-2.16^{+0.70}_{-0.52}$ & $-1.85^{+0.49}_{-0.58}$ \\
		Deep atmospheric $\text{H}_2\text{O}$ mole fraction in $\log_{10}$ & Dayside & $\mathcal{U}(-12.0,-1.0)$ & $-8.44^{+2.29}_{-2.21}$ & $-7.83^{+2.63}_{-2.49}$ \\
		$\text{H}_2\text{O}$'s knee pressure in $\log_{10}(\text{bar})$ & Dayside & $\mathcal{U}(-6.0,-1.0)$ & $-3.05^{+1.34}_{-1.76}$ & $-3.06^{+1.27}_{-1.70}$ \\
		$\text{H}_2\text{O}$'s power-law index & Dayside & $\mathcal{U}(0,3.0)$ & $1.45^{+0.91}_{-0.89}$ & $1.60^{+0.84}_{-0.93}$ \\
		$\text{CO}_2$'s mole fraction in $\log_{10}$ & Dayside & $\mathcal{U}(-12.0,-1.0)$ & $-8.64^{+1.82}_{-1.99}$ & $-8.20^{+2.11}_{-2.35}$ \\
		$\text{H}^-$'s mole fraction in $\log_{10}$ & Dayside & $\mathcal{U}(-13.0,-1.0)$ & $-9.09^{+2.38}_{-2.44}$ & $-9.01^{+2.11}_{-2.12}$ \\
		$\text{H}_2$'s knee pressure in $\log_{10}(\text{bar})$ & Dayside & $\mathcal{U}(-6.0,-1.0)$ & $-3.86^{+1.53}_{-1.26}$ & $-4.18^{+1.54}_{-1.10}$ \\
		$\text{H}_2$'s power-law index & Dayside & $\mathcal{U}(0,2.0)$ & $0.97^{+0.61}_{-0.61}$ & $0.92^{+0.58}_{-0.57}$ \\
		$\kappa_{\mathrm{IR}}^{(\mathrm{night})}$ in $\log_{10}$ (see Equation \ref{eq:pt_night}) & Nightside & $\mathcal{U}(-4.0,3.0)$ & $-0.50^{+2.09}_{-2.01}$ & $-0.52^{+2.10}_{-2.02}$ \\
		$\text{H}_2\text{O}$ mole fraction in $\log_{10}$ & Nightside & $\mathcal{U}(-12.0,-1.0)$ & $-3.13^{+1.33}_{-1.32}$ & $-2.42^{+0.69}_{-0.99}$ \\
		$\text{CO}_2$ mole fraction in $\log_{10}$ & Nightside & $\mathcal{U}(-12.0,-1.0)$ & $-8.80^{+1.64}_{-1.89}$ & $-9.02^{+1.92}_{-1.54}$ \\ \hline
	\end{tabular}
    \tablecomments{Uncertainties reported here are $1\sigma$. $\mathcal{U}$ is a uniform prior with the lower and upper edges given in parentheses. PC and TR stand for the retrieval on the phase-curve transmission spectrum and the retrieval on the transit-only transmission spectrum, respectively.}
\end{table*}

\begin{figure*}[htbp!]
    \centering
	\includegraphics[width=\hsize]{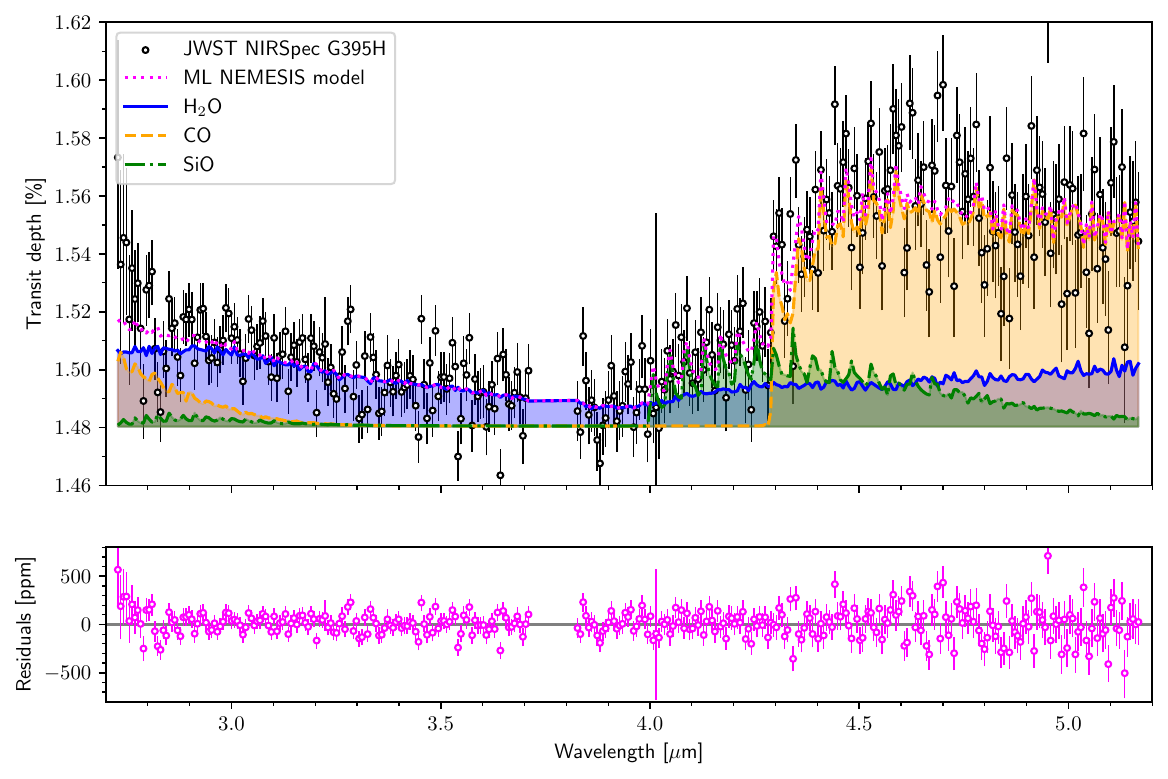}
	\caption{Contributions of the different molecular absorbers to the \texttt{NEMESIS} model spectrum. \textit{Top panel:} the black circles represent the transmission spectrum derived from the phase-curve analysis. The dotted magenta line shows the ML model and the solid blue, dashed orange, and dashed-dotted green lines show model spectra with all molecular opacities except for the opacity due to H$_2$O, CO and SiO set to zero, respectively. \textit{Bottom panel:} magenta circles depict the data subtracted by the ML \texttt{NEMESIS} model with the data's error bars propagated onto the residuals. A horizontal gray line at zero was added to guide the eye.}
	\label{fig:contributions}
\end{figure*}

To determine which molecules in WASP-121\,b's atmosphere contribute significant opacity to the transmission spectrum, we set the opacity of all molecular absorbers apart from one molecule at a time to zero in the ML model of the retrieval on the phase-curve transmission spectrum and compared the resulting synthetic spectra to the observations (Figure \ref{fig:contributions}). The dominating opacity sources in the NIRSpec G395H bandpass are
\begin{enumerate}
    \item H$_2$O, which contributes opacity throughout the entire wavelength range and creates the downward slope of the transmission spectrum between $\lambda\sim 2.7\,\mu$m and $\lambda\sim 3.7\,\mu$m;
    \item SiO which is the strongest opacity source from $\lambda\sim 4.0\,\mu$m to $\lambda\sim 4.3\,\mu$m; and
    \item CO, which dominates the opacities at $\lambda\gtrsim 4.3\,\mu$m and is responsible for the rise in transit depth at $\lambda\sim 4.6\,\mu$m.
\end{enumerate}
To quantify the statistical significance of SiO in these observations, we performed retrievals in the same setup as before, but without SiO in the atmosphere, and compared the Bayesian evidence of models including and excluding SiO. In the retrieval on the phase-curve transmission spectrum and the retrieval on the transit-only transmission spectrum, respectively, this delivered detection significances of $5.2\,\sigma$ and $2.4\,\sigma$ in favor of the model including SiO.

In addition to our \texttt{NEMESIS} retrieval that splits the planet's atmosphere along the terminator, we ran a simplified \texttt{NEMESIS} retrieval that does not include heterogeneities between the dayside and nightside of the planet to infer the Bayesian evidence for the new model. For consistency with the approach splitting the atmosphere into dayside and nightside, we kept the power-law profile parameterizations for H$_2$O and H$_2$ as well as a constant abundance for H$^-$ and used the three-parameter P-T profile parameterization that was used for the dayside in the more complex model. By comparing the Bayesian evidence between the retrievals on the phase-curve transmission spectrum, we derived a statistical significance of $1.9\,\sigma$ in favor of the atmospheric model that includes dayside and nightside heterogeneities.

\subsection{\texttt{ATMO} (Equilibrium Chemistry)}
We performed a chemical equilibrium retrieval with \texttt{ATMO} \citep{Amundsen2014A&A...564A..59A, Tremblin2015ApJ...804L..17T, Tremblin2016ApJ...817L..19T, Tremblin2017ApJ...841...30T, Drummond2016A&A...594A..69D,Goyal2018MNRAS.474.5158G,Sing2024Natur.630..831S}, a one-dimensional/two-dimensional radiative–convective equilibrium model for planetary atmospheres that has previously been used to interpret WASP-121\,b's emission and transmission spectra observed with HST \citep{Evans17,Evans18,MikalEvans20,MikalEvans22}. The chemical abundances in equilibrium were determined `on-the-fly' by minimizing the Gibbs free energy following the method of \cite{gordon1994computer}, and using the thermochemical data from \cite{mcbride1993coefficients} for 175 neutral, 9 ionic, and 93 condensate species. The molecular opacities of H$_2$O, CO, and SiO were taken from \cite{Barber2006MNRAS.368.1087B}, \cite{ROTHMAN20102139} and \cite{barton2013}, respectively and collision-induced absorption caused by H$_2$-H$_2$ and H$_2$-He collisions was modeled according to \cite{hitran_cia}. Solar elemental abundances were set from \cite{Asplund2009ARA&A..47..481A} and \cite{Caffau2011SoPh..268..255C}. The chemistry was fully flexible for any mix of input elemental abundances, and we set the oxygen, carbon and silicon abundances as free parameters relative to the solar abundances in the retrieval along with a scaled solar metallicity parameter for the remaining elements. The chemistry allowed for the depletion of gas-phase species due to condensation as well as thermal ionization and dissociation. We adopted a `Guillot' P-T profile \citep{guillot10} using two optical channels and one IR channel \citep{Line2012ApJ...749...93L,line13} with five fit parameters. The planetary radius is also free to fit along with a gray cloud. A nested sampling statistical algorithm \citep{feroz2008,feroz2009} was used to fit the parameters to the G395H data.

\begin{figure}[htbp!]
    \centering
	\includegraphics[width=\hsize]{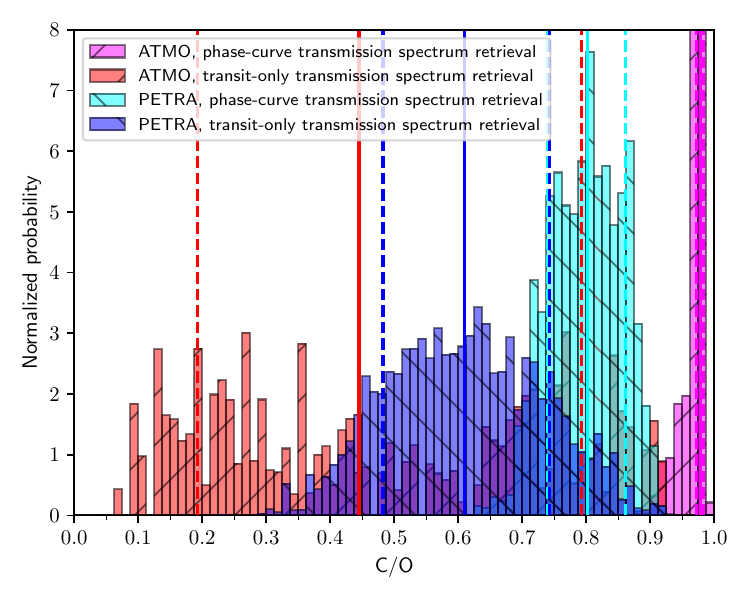}
	\caption{Posteriors for WASP-121\,b's C/O inferred from the retrievals enforcing equilibrium chemistry. The solid vertical lines depict the posteriors' medians and the dashed vertical lines show the edges of the $1\sigma$ intervals. Two bars of the posteriors of the ATMO retrieval on the phase-curve transmission spectrum exceed the y-axis: their normalized probability values are $\sim 14.5$ and $\sim 59.2$ from left to right.}
	\label{fig:CtoO}
\end{figure}

All fit parameters' posteriors of the fit to both versions of the transmission spectrum are listed in Table \ref{tab:atmo} and the model's best fits with the corresponding $1\,\sigma$ intervals are presented in Figure \ref{fig:retrieval-spectra} along with the data. For the 349 data points, we found a minimum $\chi^2_\nu=1.32$ in the retrieval on the phase-curve transmission spectrum and a minimum $\chi^2_\nu=1.13$ in the retrieval on the transit-only transmission spectrum. Oxygen and carbon are both constrained to be significantly supersolar in both retrievals, while silicon is found to be supersolar in the retrieval on the phase-curve transmission spectrum, but either subsolar, solar or supersolar in the retrieval on the transit-only transmission spectrum. The P-T profile is inverted in both retrievals with temperatures rising starting at $p\sim 0.1$\,bar in the retrieval on the phase-curve transmission spectrum (see Figure \ref{fig:retrieval-atmospheres_phase-curve}) and at $p\sim 1$\,bar in the retrieval on the transit-only transmission spectrum (see Figure \ref{fig:retrieval-atmospheres_transit-only}). The retrieval on the phase-curve transmission spectrum finds a significantly supersolar C/O of $0.978^{+0.004}_{-0.006}$, while the retrieval on the transit-only transmission spectrum finds $0.448^{+0.34}_{-0.26}$, which lies close to solar values though both high (0.9) and low (0.1) values are consistent at the 2$\,\sigma$ level (see Figure \ref{fig:CtoO}).
\begin{table*}[htbp!]
	\caption{Inputs and results of the \texttt{ATMO} retrieval.}
	\label{tab:atmo}
	\centering
    \begin{tabular}{l c r r}
		\hline\hline
        & & \multicolumn{2}{c}{\textbf{Posteriors}} \\
		\textbf{Parameter} & \textbf{Priors} & \multicolumn{1}{c}{\textbf{PC}} & \multicolumn{1}{c}{\textbf{TR}} \\ \hline
		$\alpha$ (see \citealp{Line2012ApJ...749...93L,line13}) & $\mathcal{U}(0,1)$ & $0.54^{+0.13}_{-0.07}$ & $0.41^{+0.17}_{-0.22}$ \\
		$\beta$ (see \citealp{Line2012ApJ...749...93L,line13}) & $\mathcal{U}(0,1)$ & $0.84^{+0.02}_{-0.03}$ & $0.88^{+0.06}_{-0.08}$ \\
		$\kappa_{\mathrm{IR}}$ in $\log_{10}$ (see \citealp{guillot10,Line2012ApJ...749...93L,line13}) & $\mathcal{U}(-4.0,0.5)$ & $-2.45^{+0.06}_{-0.07}$ & $-3.18^{+0.34}_{-0.45}$ \\
		$\gamma_1$ in $\log_{10}$ (see \citealp{guillot10,Line2012ApJ...749...93L,line13}) & $\mathcal{U}(-0.5,3.5)$ & $0.39^{+0.40}_{-0.38}$ & $0.92^{+0.25}_{-0.22}$ \\
		$\gamma_2$ in $\log_{10}$ (see \citealp{guillot10,Line2012ApJ...749...93L,line13})  & $\mathcal{U}(-1.5,3.5)$  & $2.57^{+0.07}_{-0.12}$ & $0.25^{+0.57}_{-0.63}$ \\
        Grey cloud deck pressure in $\ln$ & $\mathcal{U}(-10,10)$ & $-3.16^{+0.66}_{-0.33}$ & $-1.22^{+1.26}_{-1.51}$ \\
		WASP-121\,b's radius relative to Jupiter's radius & $\mathcal{U}(1.69,1.9)$ & $1.79^{+0.00}_{-0.00}$ & $1.76^{+0.01}_{-0.01}$ \\
		Elemental abundances other than H, He, C, O, and Si relative to solar in $\log_{10}$ &$\mathcal{U}(-2.8,2.8)$  & $-0.65^{+0.15}_{-0.14}$ & $0.24^{+0.88}_{-1.10}$ \\
		Carbon abundance relative to solar abundance in $\log_{10}$ & $\mathcal{U}(-1,2)$ & $1.86^{+0.06}_{-0.08}$ & $1.37^{+0.19}_{-0.19}$ \\
		Oxygen abundance relative to solar abundance in $\log_{10}$ & $\mathcal{U}(-1,2)$ & $1.16^{+0.06}_{-0.08}$ & $1.46^{+0.25}_{-0.15}$ \\
		Silicon abundance relative to solar abundance in $\log_{10}$ & $\mathcal{U}(-1,2)$ & $1.17^{+0.11}_{-0.14}$ & $0.45^{+0.31}_{-0.52}$ \\\hline
	\end{tabular}
    \tablecomments{Uncertainties reported here are $1\sigma$. $\mathcal{U}$ is a uniform prior with the lower and upper edges given in parentheses. PC and TR stand for the retrieval on the phase-curve transmission spectrum and the retrieval on the transit-only transmission spectrum, respectively.}
\end{table*}

\subsection{\texttt{PETRA} (Equilibrium Chemistry)} \label{subsec:petra_eq}
We used the \texttt{PETRA} retrieval framework \citep{petra} in a similar setup to \texttt{ATMO} in order to explore the impact of each framework's model assumptions. \texttt{PETRA} uses the \texttt{PHOENIX} atmosphere model \citep{hauschildt99,barman01} as a forward model in a Differential Evolution Markov Chain (DEMC) statistical framework \citep{terbraak06,terbraak08}. We first performed retrievals with the oxygen-to-hydrogen ratio (O/H) relative to the solar ratio and the C/O as free parameters with chemical abundances governed by chemical equilibrium. Also in line with \texttt{ATMO}, we used the same five-parameter P-T profile parameterization \citep{guillot10,Line2012ApJ...749...93L,line13}. We also included a free parameter for the reference radius of the planet and a gray cloud top pressure, below which the atmosphere becomes optically thick at all wavelengths. We took the molecular opacities for H$_2$O, CO and SiO from \cite{Barber2006MNRAS.368.1087B}, \cite{goorvitch94} and \cite{kurucz94}, respectively, and calculated H$_2$-H$_2$ and He-H$_2$ collision-induced absorption according to \cite{borysow98}.

The priors and posteriors of the fit parameters are listed in Table \ref{tab:petra-eq} for the retrievals on both data reductions' transmission spectra and show good agreement between both retrievals within the uncertainties. For the phase-curve transmission spectrum, \texttt{PETRA} returns $\chi_\nu^2=1.36$, while the retrieval on the transit-only transmission spectrum delivers $\chi_\nu^2=1.01$ (see Figure \ref{fig:retrieval-spectra}). For both versions of the transmission spectrum, our retrievals find an atmosphere with a subsolar O/H; however, the results differ for the C/O, which is tightly constrained to be supersolar in the retrieval on the phase-curve transmission spectrum and loosely constrained to be approximately solar to supersolar in the retrieval on the transit-only transmission spectrum (see Figure \ref{fig:CtoO}). The temperature profile exhibits a strong inversion beginning at about $10$\,mbar in both retrievals (see Figures \ref{fig:retrieval-atmospheres_phase-curve} and \ref{fig:retrieval-atmospheres_transit-only}), consistent with predictions from self-consistent one-dimensional radiative-convective equilibrium models \citep{lothringer18}.
\begin{table*}[htbp!]
	\caption{Results of the \texttt{PETRA} retrieval prescribing chemical equilibrium in the atmosphere.}
	\label{tab:petra-eq}
	\centering
	\begin{tabular}{l c r r}
		\hline\hline
        & & \multicolumn{2}{c}{\textbf{Posteriors}}\\
		\textbf{Parameter} & \textbf{Priors} & \multicolumn{1}{c}{\textbf{PC}} & \multicolumn{1}{c}{\textbf{TR}} \\ \hline
		$\alpha$ (see \citealp{Line2012ApJ...749...93L,line13}) & $\mathcal{U}(0,1)$ & $0.44^{+0.31}_{-0.26}$ & $0.40^{+0.26}_{-0.24}$ \\
		$\beta$ (see \citealp{Line2012ApJ...749...93L,line13}) & $\mathcal{U}(0.3,1.4)$ & $0.60^{+0.08}_{-0.08}$ & $0.64^{+0.22}_{-0.16}$ \\
		$\gamma_1$ in $\log_{10}$ (see \citealp{guillot10,Line2012ApJ...749...93L,line13}) & $\mathcal{U}(-2,4)$ & $1.92^{+0.31}_{-0.28}$ & $1.95^{+0.67}_{-0.65}$ \\
		$\gamma_2$ in $\log_{10}$ (see \citealp{guillot10,Line2012ApJ...749...93L,line13}) &  $\mathcal{U}(-2,4)$ & $1.86^{+0.34}_{-0.32}$ & $0.88^{+1.08}_{-0.81}$ \\
		$\kappa_{\mathrm{IR}}$ in $\log_{10}$ (see \citealp{guillot10,Line2012ApJ...749...93L,line13}) & $\mathcal{U}(-4,1.5)$ & $-2.73^{+0.27}_{-0.30}$ & $-2.50^{+0.51}_{-0.57}$ \\
		Cloud top pressure ($\log_{10}$ barye) & $\mathcal{U}(-3,10)$ & $5.93^{+1.44}_{-0.67}$ & $7.38^{+1.57}_{-1.34}$ \\
		WASP-121\,b's radius in $10^{10}$\,cm & $\mathcal{U}(0.1,2.5)$  & $1.195^{+0.004}_{-0.005}$ & $1.183^{+0.007}_{-0.009}$ \\
		O/H relative to solar in $\log_{10}$ & $\mathcal{U}(-2.5,1.0)$ & $-0.60^{+0.17}_{-0.15}$ & $-0.48^{+0.36}_{-0.23}$ \\
		C/O & $\mathcal{U}(0.001,100)$ & $0.80^{+0.06}_{-0.06}$ & $0.59^{+0.13}_{-0.12}$ \\ \hline
	\end{tabular}
    \tablecomments{Uncertainties reported here are $1\sigma$. $\mathcal{U}$ is a uniform prior with the lower and upper edges given in parentheses. PC and TR stand for the retrieval on the phase-curve transmission spectrum and the retrieval on the transit-only transmission spectrum, respectively.}
\end{table*}

\subsection{\texttt{PETRA} (Free Chemistry)} \label{subsec:petra_free}
To test how sensitive our interpretation of the data is to the assumption of chemical equilibrium, we also ran \texttt{PETRA} in a "free chemistry" setup, where the abundances of important molecular absorbers are allowed to vary as free parameters. In this free chemistry framework, we fit for the abundances of H$_2$O, CO and SiO with pressure-independent abundances. This allows the retrieval to fit atmospheric compositions that may deviate from chemical equilibrium, but introduces other assumptions such as a lack of thermal dissociation of these species. In contrast, the H$_2$ and H abundances were set according to chemical equilibrium to include the impact of H$_2$ dissociation without introducing additional fit parameters. We adopted the same five-parameter P-T profile, cloud parameterization, opacity information, and fit methodology of the \texttt{PETRA} retrieval enforcing chemical equilibrium (see Section \ref{subsec:petra_eq}).

The posteriors of the retrievals' fit parameters are listed in Table \ref{tab:petra_free} and agree well between the retrievals on both transmission spectra, except for the P-T profile parameters, which lead to fundamentally different P-T profiles. The retrieval on the phase-curve transmission spectrum finds temperatures that generally decrease with height (see Figure \ref{fig:retrieval-atmospheres_phase-curve}), while the retrieval on the transit-only transmission spectrum results in an inverted temperature structure (see Figure \ref{fig:retrieval-atmospheres_transit-only}). The inversion is of similar magnitude and temperature to the \texttt{PETRA} chemical equilibrium retrieval, but occurs at a lower pressure ($\sim1$\,mbar).
\begin{table*}[htbp!]
	\caption{Inputs and results for the \texttt{PETRA} retrieval employing free chemical abundances.}
	\label{tab:petra_free}
	\centering
	\begin{tabular}{l c r r}
		\hline\hline
        & & \multicolumn{2}{c}{\textbf{Posteriors}} \\
		\textbf{Parameter} & \multicolumn{1}{c}{\textbf{Prior}} & \multicolumn{1}{c}{\textbf{PC}} & \multicolumn{1}{c}{\textbf{TR}} \\ \hline
		$\alpha$ (see \citealp{Line2012ApJ...749...93L,line13}) & $\mathcal{U}(0,1)$ & $0.80^{+0.14}_{-0.19}$ & $0.41^{+0.27}_{-0.23}$ \\
		$\beta$ (see \citealp{Line2012ApJ...749...93L,line13}) & $\mathcal{U}(0.25,1.75)$ & $0.72^{+0.09}_{-0.09}$ & $0.84^{+0.12}_{-0.12}$ \\
		$\gamma_1$ in $\log_{10}$ (see \citealp{guillot10,Line2012ApJ...749...93L,line13}) & $\mathcal{U}(-2,4)$ & $0.26^{+0.68}_{-0.63}$ & $1.47^{+0.67}_{-0.65}$ \\
		$\gamma_2$ in $\log_{10}$ (see \citealp{guillot10,Line2012ApJ...749...93L,line13}) & $\mathcal{U}(-2,4)$ & $-0.99^{+0.34}_{-0.36}$ & $0.50^{+1.18}_{-1.05}$ \\
		$\kappa_{\mathrm{IR}}$ in $\log_{10}$ (see \citealp{guillot10,Line2012ApJ...749...93L,line13}) & $\mathcal{U}(-4,1.5)$ & $-0.17^{+0.48}_{-0.53}$ & $-0.63^{+0.55}_{-0.74}$ \\
		Cloud top pressure ($\log_{10}$ Ba) & $\mathcal{U}(-3,15)$ & $10.02^{+2.85}_{-2.94}$  & $4.26^{+0.48}_{-0.43}$ \\
		WASP-121\,b's radius in $10^{10}$\,cm & $\mathcal{U}(0.1,2.5)$ & $1.231^{+0.008}_{-0.009}$ & $1.218^{+0.006}_{-0.006}$ \\
		$\text{H}_2\text{O}$ mole fraction in $\log_{10}$ & $\mathcal{U}(-15,-2)$ & $-5.74^{+0.12}_{-0.11}$ & $-4.95^{+0.58}_{-0.46}$ \\
		$\text{CO}$ mole fraction in $\log_{10}$ & $\mathcal{U}(-15,-2)$ & $-2.34^{+0.24}_{-0.34}$ & $-2.90^{+0.59}_{-0.67}$ \\
		$\text{SiO}$ mole fraction in $\log_{10}$ & $\mathcal{U}(-15,-2)$ & $-4.60^{+0.16}_{-0.17}$ & $-3.87^{+0.62}_{-0.52}$ \\ \hline
	\end{tabular}
    \tablecomments{Uncertainties reported here are $1\sigma$. $\mathcal{U}$ is a uniform prior with the lower and upper edges given in parentheses. PC and TR stand for the retrieval on the phase-curve transmission spectrum and the retrieval on the transit-only transmission spectrum, respectively.}
\end{table*}

The phase-curve transmission spectrum is fit with a $\chi^2_\nu=1.47$ and the fit to the transit-only transmission spectrum results in $\chi^2_\nu=1.20$ (see Figure \ref{fig:retrieval-spectra}). In both retrievals, the CO abundance posterior is limited by running up against the upper edge of the uniform prior at $1.0\times 10^{-2}$ (see Table \ref{tab:petra_free}), which might have an impact on the goodness of fit. In the retrievals on both versions of the transmission spectrum, the H$_2$O and SiO abundances are generally consistent with the \texttt{PETRA} retrievals prescribing chemical equilibrium; however, the CO abundances of the free chemistry retrievals are both higher than those in the chemical equilibrium retrievals (see Figures \ref{fig:retrieval-atmospheres_phase-curve} and \ref{fig:retrieval-atmospheres_transit-only}). To test whether including SiO in the model is justified, we also ran the same free chemistry retrieval setup without SiO and calculated the Bayesian information criterion (BIC) between the two different models for both the retrieval on the phase-curve transmission spectrum and the retrieval on the transit-only transmission spectrum. This resulted in a $\Delta\text{BIC}=2.0$ in the retrieval on the phase-curve transmission spectrum and a $\Delta\text{BIC}=8.9$ in the retrieval on the transit-only transmission spectrum in favor of the model including SiO.

\subsection{GCM comparison} \label{subsec:gcm}
WASP-121\,b's atmosphere is expected to be extremely inhomogeneous due to the drastic differences in temperatures, and thus the thermal dissociation, between the dayside and the nightside (see Section \ref{sec:introduction}). As these effects are intrinsically three-dimensional, GCMs can yield important insight into the physics of the atmosphere when comparing their simulated transmission spectra to observations.

\begin{figure*}[htbp!]
    \centering
	\includegraphics[width=\hsize]{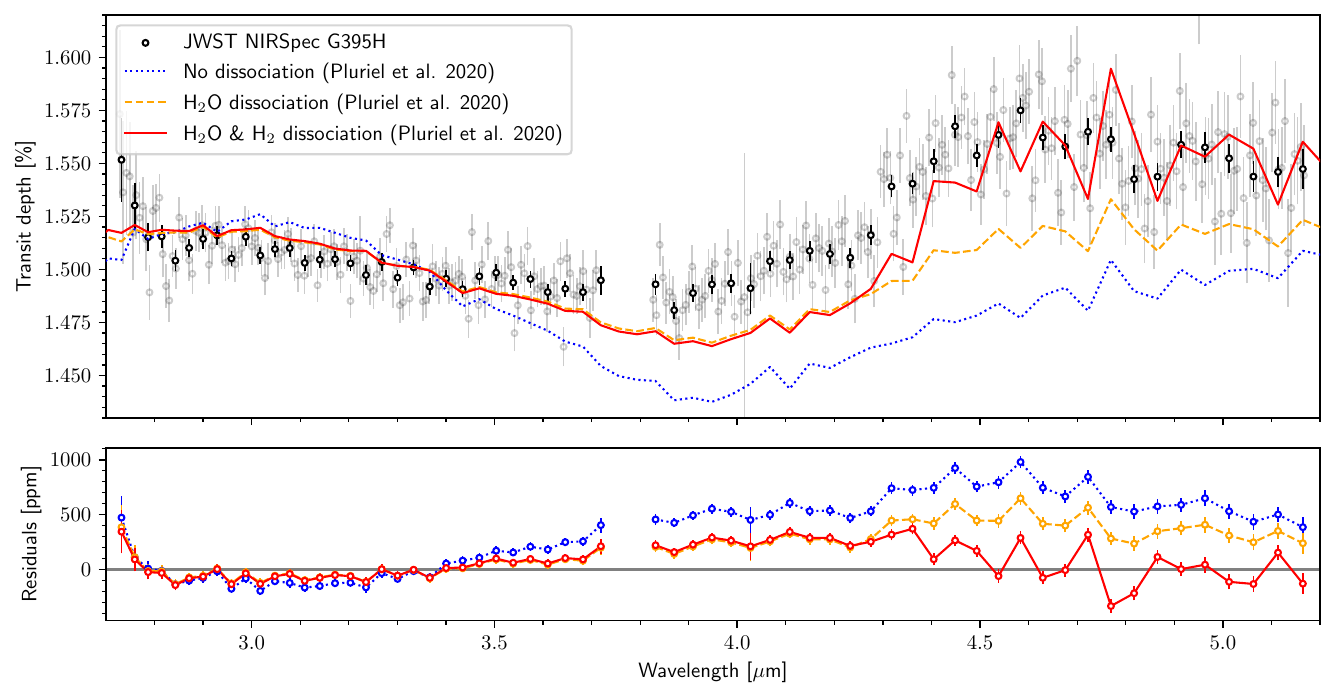}
	\caption{WASP-121\,b's transmission spectrum obtained from the phase-curve analysis at a spectral resolution of $R\sim 600$ and the GCM transmission spectra presented by \cite{pluriel20}. \textit{Upper panel:} the gray circles show the original data and the black circles show the data binned in wavelength. The model spectra presented by \cite{pluriel20} with varying degrees of molecular dissociation are plotted using lines. A wavelength-independent offset was added to the model spectra to fit the mean of the data between $2.9\,\mu$m and $3.7\,\mu$m. \textit{Lower panel:} the residuals between the model spectra with an added constant and the data are depicted using colored circles with the error bars of the data propagated onto the residuals. A horizontal gray line at zero was added to guide the eye.}
	\label{fig:gcm}
\end{figure*}

Figure \ref{fig:gcm} shows WASP-121\,b's transmission spectrum derived from the phase-curve analysis of the NIRSpec G395H observation together with model spectra derived by \cite{pluriel20} from the GCM of \cite{parmentier18}. The GCM of \cite{parmentier18} did not include the effects of thermal dissociation on the atmosphere's mean molecular weight and latent heat, so \cite{pluriel20} postprocessed the effect of H$_2$O and H$_2$ dissociation onto the GCM output to demonstrate its impact on the transmission spectrum. When no dissociation is taken into account (blue dotted line), the model spectrum poorly represents the data, since the slope of the H$_2$O feature on the NRS1 detector is too steep and the amplitude of the CO feature on the NRS2 detector is too small. Including H$_2$O dissociation (orange dashed line) flattens the slope of the H$_2$O feature, since the dissociation reduces the H$_2$O abundance on the dayside. The resulting absorption band thus originates from the nightside which creates smaller variations in transit depth due to its lower temperatures and higher mean molecular weight compared to the dayside. Further incorporating H$_2$ dissociation into the model (red solid line) mostly acts to increase the amplitude of the CO absorption band, because the resulting decrease in mean molecular weight inflates the atmospheric scale height of the dayside, leading to an increase in the strength of the absorption features of any molecule that is present on the dayside. As CO, unlike H$_2$O, is abundant on the dayside, including H$_2$ dissociation changes the observed CO band, but not the H$_2$O band.

\cite{pluriel20}'s model spectra only include CO and H$_2$O, but not SiO opacities, leading to a substantial underestimation of the transit depth between $4.0\,\mu$m and $4.3\,\mu$m in all models. However, it is notable that the agreement between the data and models progressively improves as more realistic three-dimensional effects, namely the effects of the thermal dissociation of H$_2$O and H$_2$ on the mean molecular weight and latent heat of the atmosphere, are added.

\section{Discussion}
\subsection{Thermal dissociation}
Thermal dissociation and molecular recombination strongly depend on temperature, and thus, in a tidally locked planet such as WASP-121\,b, they are variable in latitude, longitude, and altitude (see, e.g., \citealp{bell18,parmentier18}). Fully grasping their effects on the transmission spectrum thus requires a three-dimensional model. The GCM simulations presented by \cite{pluriel20} showcase the importance of both H$_2$ and H$_2$O dissociation on WASP-121\,b's dayside as including H$_2$O dissociation is needed to fit the slope of the transmission spectrum between $2.7$ and $3.7\,\mu$m and the dissociation of H$_2$ inflates the CO absorption feature at wavelengths longer than $4.3\,\mu$m to the observed levels (see Figure \ref{fig:gcm}). These findings suggest that including changes of the mean molecular weight and latent heat induced by thermal dissociation in the model leads to a more accurate representation of the planet's atmospheric physics.

The role of thermal dissociation in our \texttt{NEMESIS}, \texttt{ATMO} and \texttt{PETRA} models that are all not fully three-dimensional is less evident. \texttt{NEMESIS} is our only model that is capable of grasping the thermochemical differences between the two hemispheres of WASP-121\,b as we divided the ray across the atmosphere into dayside and nightside integrals and fit for different thermochemical conditions in both hemispheres. Our \texttt{NEMESIS} retrievals favor a decrease of both the H$_2$ and H$_2$O abundances in the upper dayside atmosphere (see Figures \ref{fig:retrieval-atmospheres_phase-curve} and \ref{fig:retrieval-atmospheres_transit-only}) as well as a higher H$_2$O abundance on the nightside than on the dayside in the lower atmosphere ($p\gtrsim 0.1$\,bar) consistent with thermal dissociation. However, the decreases with height of both the H$_2$ abundance ($\alpha_{\text{H}_2}=0.97^{+0.61}_{-0.61}$ and $\alpha_{\text{H}_2}=0.92^{+0.58}_{-0.57}$ in the retrievals on the phase-curve and transit-only transmission spectra, respectively) and the H$_2$O abundance ($\alpha_{\text{H}_2\text{O}}=1.45^{+0.91}_{-0.89}$ and $\alpha_{\text{H}_2\text{O}}=1.60^{+0.84}_{-0.93}$ in the retrievals on the phase-curve and transit-only transmission spectra, respectively) are consistent with zero within $2\,\sigma$. The differences between the deep atmospheric water abundances of the dayside ($-8.44^{+2.29}_{-2.21}$ and $-7.83^{+2.63}_{-2.49}$ in the retrievals on the phase-curve and transit-only transmission spectra, respectively) and the nightside  ($-3.13^{+1.33}_{-1.32}$ and $-2.42^{+0.69}_{-0.99}$ in the retrievals on the phase-curve and transit-only transmission spectra, respectively) are $2.0\sigma$ in the retrieval on the phase-curve transmission spectrum and $1.9\sigma$ in the retrieval on the transit-only transmission spectrum. Thus, these differences are also not significantly different from zero. However, a lower dayside H$_2$O abundance compared to the nightside is in agreement with GCM simulations \citep{parmentier18}, suggesting that the different H$_2$O abundances between the hemispheres in our retrievals might be driven by thermal dissociation. Similar to our results, recent Gemini-S/IGRINS observations of WASP-121\,b suggest a higher H$_2$O abundance on the dayside than on the nightside \citep{wardenier24} and its phase-curve observed using HST \citep{MikalEvans22} points toward a difference of many orders of magnitude in the dayside and nightside H$_2$O abundances in the part of the atmosphere that is typically observed in transmission spectroscopy ($p\sim 10^{-3}$\,bar). \cite{MikalEvans22} inferred very similar dayside and nightside H$_2$O abundances in the lower atmosphere ($p\gtrsim 10^{-2}$\,bar); however, as observations of thermal emission of hot Jupiters are generally sensitive to higher pressure-ranges compared to transit observations, that part of the atmosphere is probably unconstrained by our analysis of WASP-121\,b's transmission spectrum.

The decrease of the quality of fit when moving from a chemical equilibrium model including thermal dissociation to one with free chemistry but with pressure-independent abundances and thus no thermal dissociation within the \texttt{PETRA} retrieval framework might be a potential hint toward the importance of thermal dissociation. Despite having one more free parameter, the retrievals with free chemistry deliver $\chi^2_\nu=1.47$ for the phase-curve transmission spectrum and $\chi^2_\nu=1.20$ for the transit-only transmission spectrum, while the retrievals enforcing chemical equilibrium return $\chi_\nu^2=1.36$ for the phase-curve transmission spectrum and $\chi_\nu^2=1.01$ for the transit-only transmission spectrum. However, these differences in $\chi_\nu^2$ are not conclusive evidence for thermal dissociation and might also be caused by the free chemistry retrievals being limited by running into the upper edge of the uniform prior on the CO abundance (see Section \ref{subsec:petra_free}).

\subsection{Molecular abundances} \label{subsec:abundances}
Dichotomies between different molecules' global distributions probably leave an impact on transmission spectra (see Section \ref{sec:introduction}). The much higher temperatures on the dayside, compounded by the dissociation of H$_2$ into atomic hydrogen, act to increase the scale height over that on the nightside, leading to an inflation of the absorption features of any molecule that is abundant on the dayside, such as CO (see Figure \ref{fig:gcm}). This potentially gives rise to a negative correlation between the amount of H$_2$ dissociation and the CO abundances inferred from atmospheric models.

We potentially observe this degeneracy in our atmospheric models. The \texttt{NEMESIS} model, which finds the highest CO abundance among our models, also finds less H$_2$ dissociation than the \texttt{ATMO} and \texttt{PETRA} equilibrium chemistry models, since \texttt{NEMESIS}'s H$_2$ abundances are higher in all parts of the atmosphere than the other two models' H$_2$ abundances (see Figures \ref{fig:retrieval-atmospheres_phase-curve} and \ref{fig:retrieval-atmospheres_transit-only}). Therefore, the CO abundance inferred using \texttt{NEMESIS} might be partly driven by less H$_2$ dissociation. The only model that finds less H$_2$ dissociation than \texttt{NEMESIS} is \texttt{PETRA} in the free chemistry setup. However, the CO abundance posteriors of the free chemistry \texttt{PETRA} retrievals are limited by the upper edge of the adopted uniform prior at $1\%$ (see Table \ref{tab:petra_free}), which prevents the posterior from reaching similar CO levels to those of the \texttt{NEMESIS} retrieval. Within the \texttt{PETRA} framework, the CO abundance of the free chemistry retrieval, which includes less H$_2$ dissociation than the chemical equilibrium model, is higher than the chemical equilibrium model's CO abundance. Thus, the \texttt{PETRA} retrievals follow the trend of including higher CO abundances with less H$_2$ dissociation. \texttt{ATMO} finds lower CO abundances than \texttt{NEMESIS}, but higher CO abundances than \texttt{PETRA}. However, in the retrieval on the phase-curve transmission spectrum (see Figure \ref{fig:retrieval-atmospheres_phase-curve}), the \texttt{ATMO} model reaches temperatures high enough to dissociate CO, which probably complicates the picture between H$_2$ dissociation and CO abundances. Therefore, all CO abundances we inferred from the different atmospheric models might be biased and we cannot derive a quantitative constraint.

The H$_2$O abundances we derived using \texttt{ATMO} and \texttt{PETRA} agree around $0.1-1$\,mbar (see Figures \ref{fig:retrieval-atmospheres_phase-curve} and \ref{fig:retrieval-atmospheres_transit-only}), the part of the atmosphere that is typically observed in transmission spectroscopy. The dayside and nightside H$_2$O abundance profiles from \texttt{NEMESIS} deliver lower and higher abundances, respectively, than the other models' abundance profiles in the observable part of the atmosphere. Thus, the inferred abundances are to first order compatible with dissociation on the dayside and with recombination on the nightside. This demonstrates that modeling the thermochemical conditions of both hemispheres is crucial for constraining the planet's bulk H$_2$O content as the bulk inventory is approximated more closely by the nightside abundance that is not modified by dissociation. 

For SiO, our models find consistent abundances when retrieving on the transit-only transmission spectrum (see Figure \ref{fig:retrieval-atmospheres_transit-only}) and abundances spanning from $\sim 2\times 10^{-5}$ to $\sim 10^{-3}$ when fitting the phase-curve transmission spectrum (see Figure \ref{fig:retrieval-atmospheres_phase-curve}). Our \texttt{ATMO} and \texttt{PETRA} retrievals enforcing equilibrium chemistry suggest that SiO is more susceptible to thermal dissociation than CO, but more stable against it than H$_2$O, as the decrease of its abundances starts at lower pressures and follows smaller gradients (see Figures \ref{fig:retrieval-atmospheres_phase-curve} and \ref{fig:retrieval-atmospheres_transit-only}). Thus, the degree to which the SiO abundances are impacted by a degeneracy with H$_2$ dissociation and the degree of reliability of our inferred abundances remain unclear. However, the fact that both \texttt{ATMO} and \texttt{PETRA} find observable amounts of SiO in equilibrium chemistry delivers a direct explanation for SiO in WASP-121\,b: SiO is a molecule whose observable presence in WASP-121\,b does not require disequilibrium processes such as photochemistry or mixing from deeper levels of the atmosphere.

\subsection{Elemental abundance ratios}
The constraints on WASP-121\,b's C/O we inferred using \texttt{ATMO} and \texttt{PETRA} are informed by absorption from CO, SiO, and H$_2$O. As these molecules are susceptible to thermal dissociation in different degrees (see Section \ref{subsec:abundances}), the three-dimensional nature of dissociation probably biases the results of this elemental abundance ratio \citep{pluriel20}.

Both \texttt{ATMO} and \texttt{PETRA} prescribing chemical equilibrium constrain WASP-121\,b's C/O to be significantly supersolar in their respective retrievals on the phase-curve transmission spectrum ($0.978^{+0.004}_{-0.006}$ in \texttt{ATMO} and $0.80^{+0.06}_{-0.06}$ in \texttt{PETRA}; see Figure \ref{fig:CtoO}). In the retrievals on the transit-only transmission spectrum, \texttt{PETRA} favors a solar to supersolar C/O ($0.59^{+0.13}_{-0.12}$), while \texttt{ATMO} finds solutions compatible with both a supersolar and a subsolar C/O ($0.45^{+0.34}_{-0.26}$). Therefore, our retrievals appear to overall favor WASP-121\,b to have a supersolar C/O. However, given the challenges in deriving the abundances of molecules whose absorption features are sensitive to the different hemispheres of the planet when applying one-dimensional models \citep{pluriel20} and the fact that our results span $\text{C}/\text{O}=0.19$ to $\text{C}/\text{O}=0.98$ in the retrieval posteriors' $1\,\sigma$ intervals, we cannot derive a quantitative constraint. Other elemental ratios such as refractory-to-volatile ratios (e.g., Si/C or Si/O) would also rely on the relative absorption strengths of SiO, CO, and H$_2$O. Thus, these ratios are probably susceptible to biases similar to the ones affecting C/O and we refrain from attempting to constrain them. 

\subsection{Identification of absorbers} \label{subsec:detections}
The challenges in constraining the abundances of absorbing molecules and elemental abundance ratios in WASP-121\,b described above do not affect our ability to identify the molecules responsible for the observed atmospheric absorption (see, e.g., \citealp{YassinJaziri24}). What drives the detection significances of molecules in the atmosphere is the model's fit to the observed transmission spectrum, which depends on the shape of the molecules' absorption features in wavelength. As these absorption features' shapes do not change significantly when moving to more complex atmospheric models, it is unlikely that any absorber's detection would disappear when moving to a more complex atmospheric model.

Since the shape of the transmission spectrum is matched closely by the H$_2$O and CO opacities in both our models (see Figure \ref{fig:contributions}) and the models presented by \cite{pluriel20} that only include CO and H$_2$O opacities (Figure \ref{fig:gcm}), both molecules' contributions to the transmission spectrum are evident. Notably, there is a clear underestimation of atmospheric absorption in the model spectra presented by \cite{pluriel20} between $\lambda\sim 4.0$ and $\lambda\sim4.3\,\mu$m (see Figure \ref{fig:gcm}), hinting at the existence of at least one more absorber in addition to H$_2$O and CO. In our \texttt{NEMESIS} model, the dominant absorber in that wavelength range is SiO, which contributes more opacity than both H$_2$O and CO (see Figure \ref{fig:contributions}). To quantify the statistical significance of SiO in these observations, we performed retrievals excluding SiO from the atmosphere in \texttt{NEMESIS} and \texttt{PETRA}, both allowing free chemistry. Comparing the Bayesian evidence of the \texttt{NEMESIS} retrievals delivers a conclusive detection significance of $5.2\sigma$ in favor of including SiO in the model in the retrieval on the phase-curve transmission spectrum and a tentative detection of $2.4\sigma$ in the retrieval on the transit-only transmission spectrum. When going from the SiO-less model to the one including SiO in \texttt{PETRA} modeling free chemical abundances, the changes in the BIC are 2.0 and 8.9 in the retrieval on the phase-curve transmission spectrum and the retrieval on the transit-only transmission spectrum, respectively, delivering moderate and strong evidence in favor of including SiO. As the phase-curve transmission spectrum has been shown to be more precise than the transit-only transmission spectrum thanks to the inclusion of all observations in the data analysis (see Section \ref{subsec:reduction-comparison}), we consider inferences from the former data reduction more reliable. The \texttt{NEMESIS} framework appears to be the most reliable retrieval for WASP-121\,b's transmission spectrum, as its results agree the most between the retrievals on both versions of the transmission spectrum (see Tables \ref{tab:nemesis}, \ref{tab:atmo}, \ref{tab:petra-eq}, and \ref{tab:petra_free}), as it allows for the expected heterogeneity between WASP-121\,b's dayside and nightside and as it delivers the lowest $\chi_\nu^2$ values for both versions of the transmission spectrum. Additionally, and unlike \texttt{PETRA}, it directly calculates the Bayesian evidence. Thus, we consider the detection significance of $5.2\sigma$ found in the \texttt{NEMESIS} retrieval on the phase-curve transmission spectrum the most reliable one.

SiO is a molecule consistent with chemical equilibrium in WASP-121\,b (see Section \ref{subsec:abundances}), providing a direct explanation of its presence in observable abundances. Additionally, it has previously been suggested as a source of the observed NUV opacity in WASP-121\,b and WASP-178\,b \citep{lothringer20,lothringer22}. Our detection of SiO is thus in line with its previously hypothesized impact on WASP-121\,b's NUV transmission spectrum.

\subsection{Light-curve systematics}
The fit to the white light curves (see Figure \ref{fig:white_lightcurves}) delivers a $\chi_\nu^2$ of $1.42$ for the simultaneous fit. The spectrophotometric light-curve fits at both presented spectral resolutions appear to be effectively free of systematic residuals (see Figure \ref{fig:allan_plot}) and the $\chi_\nu^2$ of the fits to the $R\sim 600$ light curves are mostly close to 1 (see Figure \ref{fig:spectroscopic_fit_parameters}).

However, there is one apparent feature in the residuals of the fit to the white light curve of NRS2 that is absent in the fit to the white light curve of NRS1. This feature is a collection of negative residuals between $\text{BJD}\_{\text{TDB}}=2459867.6632$ and $\text{BJD}\_{\text{TDB}}=2459867.6894$, just before egress starts (see Figure \ref{fig:white_lightcurves}, the dashed line in the lower panel of Figure \ref{fig:spectroscopic_lightcurves} and \citealp{MikalEvans23}). As these residuals are not symmetric about the transit midtime, they are most likely explained by instrumental or stellar systematics or deviations of the planet's shape from a sphere. Recent studies of transit signals of exoplanets with asymmetric terminators have shown, however, that limb asymmetries mostly affect the transit signal during ingress and egress (see, e.g., \citealp{catwoman,grant23}). Indeed, the unexplained residuals are also apparent in our \texttt{catwoman} fits to the white light curves (see Figure \ref{fig:catwoman_white}). Thus, it is unlikely that these residuals we observe are caused by asymmetric limbs. Alternatively, WASP-121\,b's rotation during the transit might create the unexplained residuals before egress in the NRS2 light curve: as the hottest and thus most extended parts of the dayside rotate into view, the absorbing cross-sectional area of the planet increases with time \citep{falco24}. Possibly, only the NRS2 but not the NRS1 light curve is sensitive to this effect, because CO absorbs in NRS2's bandpass, while the NRS1 bandpass is mostly sensitive to H$_2$O (see Figure \ref{fig:contributions}). Thus, the dayside is effectively transparent in the NRS1 light curve due to the thermal dissociation of H$_2$O, but absorbing in the NRS2 light curve, making the NRS1 light curve sensitive to the nightside only, while the NRS2 light curve probes both the dayside and the nightside.

As none of the identified residuals are apparent in each individual spectrophotometric light-curve fit (see Figure \ref{fig:spectroscopic_lightcurves}), we are confident that these residuals do not leave a significant impact on the extracted transmission spectrum. However, modeling the impact of the dayside and nightside on the white light curves might be a promising path toward understanding the residuals in our NRS2 white light curve fits.

\section{Summary and conclusions}
The JWST observations of a full phase-curve of the ultrahot Jupiter WASP-121\,b using NIRSpec G395H have delivered an observation of the planet's transmission spectrum between $\sim 2.7\,\mu$m and $\sim 5.2\,\mu$m of unprecedented quality. Analyzing the planet's full phase-curve instead of a cutout of the transit from the observations appears to reduce the contamination of the inferred transmission spectrum with the planet's nightside emission significantly. Additionally, the full phase-curve analysis delivers a more precise transmission spectrum with $\sim 40\,\%$ smaller error bars due to stronger constraints on the phase-curve shape, which is correlated with the transit depth. Both the white and spectrophotometric light curves show no evidence for asymmetric limbs in WASP-121\,b, allowing an analysis of the transmission spectrum without modeling differences between the planet's morning and evening terminators.

H$_2$O and CO are apparent in the transmission spectrum, with evidence for the existence of at least one more molecule absorbing stellar radiation between $\lambda\sim 4.0\,\mu$m and $\lambda\sim 4.3\,\mu$m. From our \texttt{NEMESIS} retrievals, we identify that absorber as SiO at a statistical significance of $5.2\sigma$. Comparative retrievals within the \texttt{ATMO} and \texttt{PETRA} frameworks that enforce chemical equilibrium show that the observed SiO feature is compatible with equilibrium chemistry in the atmosphere. Thus, not only does SiO fit the observed spectral feature, its presence in WASP-121\,b at observable quantities also does not require disequilibrium chemistry.

Previous modeling efforts simulating WASP-121\,b's transmission spectrum using a GCM have shown that changes in the mean molecular weight and latent heat induced by thermal dissociation of H$_2$O and H$_2$ modify the shapes of the absorption features of H$_2$O and CO \citep{pluriel20}. Our JWST/NIRSpec G395H observations are approximated best when including both H$_2$O and H$_2$ dissociation, demonstrating the importance of thermal dissociation when deriving the transmission spectrum from the planet's three-dimensional shape. As CO is more stable against thermal dissociation than H$_2$O, the observed CO feature is sensitive to H$_2$ dissociation on the dayside, while the H$_2$O feature is not. Therefore, quantitative constraints on these molecules' abundances require an adequate representation of the thermochemical heterogeneity between WASP-121\,b's dayside and nightside. We attempted to model this heterogeneity using the \texttt{NEMESIS} framework by splitting the ray through the atmosphere into dayside and nightside integrals. The atmospheric model inferred from the \texttt{NEMESIS} retrieval fits the observations well, but does not deliver a statistically significant improvement over a comparatively simpler model without hemispheric differences. However, the retrieval allowing chemical differences between the two hemispheres finds a higher H$_2$O abundance on the nightside than on the dayside, demonstrating the impact of hemispheric heterogeneity when attempting to constrain the planet's bulk H$_2$O inventory. From the comparison of model findings within the \texttt{NEMESIS}, \texttt{ATMO} and \texttt{PETRA} frameworks, we conclude that molecular abundances and elemental abundance ratios are prone to biases and degeneracies that might be caused by a negative correlation between H$_2$ dissociation and CO abundances. These issues, which stem from the three-dimensional nature of the planet, prevent us from constraining molecular abundances and elemental abundance ratios from the observed transmission spectrum with the analyses we present here.

Constraining the abundance of SiO in WASP-121\,b as well as the abundances of H$_2$O and CO would open up new observables of the planet's chemical inventory, such as refractory-to-volatile ratios. These ratios would inform us about the amount of rocky material accreted by the planet during its formation and thus help to constrain its migration history. However, as the molecules needed to establish a picture of WASP-121\,b's elemental abundance ratios differ in their stabilities against thermal dissociation, any quantitative constraints require a careful representation of the three-dimensional effects of dissociation in the applied atmospheric model.

\begin{acknowledgments}
This work is based on observations made with the NASA/ESA/CSA James Webb Space Telescope. The data were obtained from the Mikulski Archive for Space Telescopes at the Space Telescope Science Institute, which is operated by the Association of Universities for Research in Astronomy, Inc., under NASA contract NAS 5-03127 for JWST. The specific observations analyzed can be accessed via\dataset[DOI: 10.17909/6qnn-6j23]{https://doi.org/10.17909/6qnn-6j23}. C.G. and T.M.E.-S. thank Jérémy Leconte and Tiziano Zingales for fruitful discussions and for providing the simulated transmission spectra published in \cite{pluriel20} and plotted in Figure \ref{fig:gcm} in machine-readable format. J.K.B. is supported by a Science and Technology Facilities Council Ernest Rutherford Fellowship [ST/T004479/1]. N.J.M. acknowledges support from a UKRI Future Leaders Fellowship [MR/T040866/1], a Science and Technology Facilities Council consolidated grant [ST/R000395/1] and the Leverhulme Trust through a research project grant [RPG-2020-82].
\end{acknowledgments}

\facilities{JWST.}
\software{\texttt{astropy} \citep{astropy13,astropy18,astropy22}, \texttt{ATMO} \citep{Amundsen2014A&A...564A..59A, Tremblin2015ApJ...804L..17T, Tremblin2016ApJ...817L..19T, Tremblin2017ApJ...841...30T, Drummond2016A&A...594A..69D,Goyal2018MNRAS.474.5158G}, \texttt{batman} \citep{batman}, \texttt{catwoman} \citep{catwoman,catwoman2}, \texttt{corner} \citep{corner}, \texttt{dynesty} \citep{dynesty}, \texttt{emcee} \citep{emcee}, \texttt{ExoTIC-LD} \citep{exoticld}, \texttt{FIREFLY} \citep{rustamkulov22,rustamkulov23}, \texttt{lmfit} \citep{lmfit}, \texttt{matplotlib} \citep{matplotlib}, \texttt{MultiNest} \citep{feroz2009,Feroz2019}, \texttt{NEMESIS} \citep{irwin08,lee12}, \texttt{numpy} \citep{numpy}, \texttt{pandas} \citep{pandas}, \texttt{PETRA} \citep{petra}, \texttt{PyMultiNest} \citep{Buchner2014}, \texttt{scipy} \citep{scipy}. }

\bibliography{literature}{}
\bibliographystyle{aasjournal}

\end{document}